\documentclass[prd,preprint,superscriptaddress,preprintnumbers,eqsecnum,showpacs,nofootinbib,nobibnotes]{revtex4}
\usepackage{amsfonts,bm,mathbbol}
\usepackage{graphicx}
\usepackage{pstricks}

\newcommand{\be}{\begin{equation}}
\newcommand{\bea}{\begin{eqnarray}}
\newcommand{\ee}{\end{equation}}
\newcommand{\eea}{\end{eqnarray}}
\newcommand{\sla}{\slash \hspace{-0.22cm}}

\def\s#1{{\scriptscriptstyle #1}}
\def\1eq#1{Eq.~(\ref{#1})}
\def\2eqs#1#2{Eqs.~(\ref{#1}) and~(\ref{#2})}
\def\3eqs#1#2#3{Eqs.~(\ref{#1}), (\ref{#2}) and~(\ref{#3})}
\def\4eqs#1#2#3#4{Eqs.~(\ref{#1}), (\ref{#2}), (\ref{#3}) and~(\ref{#4})}
\def\noeq#1{(\ref{#1})}

\def\fig#1{Fig.~\ref{#1}} 

\def\kint{\int_k\!}
\def\eint{\int_{\s{\mathrm{E}}}\!}
\def\diff#1{{\rm d}^{#1}}
\def\qm{{\cal M}}

\def\quark{\widehat{X}}
\def\quarkren{\widehat{X}_{\s R}}
\def\dress{{\cal Z}}
\def\quarkCP{\quark^{\s{\rm CP}}}

\def\pslash{p\hspace{-0.18cm}\slash}

\def\ie{{\it i.e.}, }

\def\bqq{\mathrm{I}\!\Gamma}

\def\n#1{({\it #1}\,)}

\def\G{\Gamma}

\begin{document}

%\title{Gluon propagator with non-perturbative quark loop effects}

\title{Unquenching the gluon propagator \\ with Schwinger-Dyson equations}
%\date{February 28, 2011}

\author{A.~C. Aguilar}
%\email{Arlene.Aguilar@ufabc.edu.br}
\affiliation{Federal University of ABC, CCNH, \\
Rua Santa Ad\'{e}lia 166, CEP 09210-170, Santo Andr\'{e}, Brazil.}

\author{D. Binosi}
%\email{binosi@ectstar.eu}
\affiliation{European Centre for Theoretical Studies in Nuclear
Physics and Related Areas (ECT*) and Fondazione Bruno Kessler, \\Villa Tambosi, Strada delle
Tabarelle 286, 
I-38123 Villazzano (TN)  Italy}

\author{J. Papavassiliou}
%\email{Joannis.Papavassiliou@uv.es}
\affiliation{\mbox{Department of Theoretical Physics and IFIC, 
University of Valencia and CSIC},
E-46100, Valencia, Spain}

\begin{abstract}

In this  article we use  the Schwinger-Dyson equations to  compute the
nonperturbative  modifications caused  to the  infrared  finite gluon
propagator (in the Landau gauge) by the inclusion of a small number of
quark families.  Our basic operating  assumption is that the main bulk
of  the  effect  stems   from  the  ``one-loop  dressed''  quark  loop
contributing to  the full gluon  self-energy. This quark loop  is then
calculated, using  as basic ingredients the full  quark propagator and
quark-gluon  vertex; for  the  quark propagator  we  use the  solution
obtained from the  quark gap equation, while for  the vertex we employ
suitable Ans\"atze, which guarantee  the transversality of the answer.
The resulting effect is included as a correction to the quenched gluon
propagator, obtained in recent  lattice simulations.  Our main finding
is that the unquenched  propagator displays a considerable suppression
in the intermediate momentum  region, which becomes more pronounced as
we increase the number of active quark families.  The influence of the
quarks on  the saturation point  of the propagator cannot  be reliably
computed within the present scheme; the general tendency appears to be
to decrease  it, suggesting a corresponding increase  in the effective
gluon mass.   The renormalization properties  of our results,  and the
uncertainties  induced  by  the  unspecified transverse  part  of  the
quark-gluon vertex,  are discussed.  Finally, the  dressing function of
the gluon propagator is compared with the available unquenched lattice
data, showing rather good agreement.

\end{abstract}


\maketitle

\section{Introduction}

In recent years considerable progress has been made in our understanding 
of various aspects of the nonperturbative dynamics of Yang-Mills theories,
through the fruitful combination of a variety of approaches and techniques~\cite{Cucchieri:2007md,Cucchieri:2007rg,Cucchieri:2009zt,
Cucchieri:2011ga,Kamleh:2007ud,
Bowman:2007du,Bogolubsky:2007ud,Bogolubsky:2009dc,Oliveira:2008uf,Oliveira:2009eh,Alkofer:2000wg,Fischer:2006ub,
Aguilar:2006gr,Binosi:2007pi,Aguilar:2008xm, Binosi:2009qm, RodriguezQuintero:2011vw,
RodriguezQuintero:2010wy, Boucaud:2010gr, Boucaud:2008gn, Boucaud:2008ji,Fischer:2008uz,Szczepaniak:2010fe,Aguilar:2004sw,Dudal:2008sp,Dudal:2010tf,Dudal:2011gd,Kondo:2011ab}.
Particularly successful in this effort has been the 
continuous interplay between lattice simulations and Schwinger-Dyson 
equations (SDEs)~\cite{Aguilar:2010zx,Aguilar:2010gm,Aguilar:2010cn,Aguilar:2011yb,Cucchieri:2011ig,Dudal:2012hb},
which has led to a firmer grasp on the infrared (IR) behavior of the fundamental 
Green's functions of QCD, such as gluon, ghost, and quark 
propagators, as well as some of the basic vertices of the theory, 
for special kinematic configurations~\cite{Skullerud:2003qu,Cucchieri:2008qm,Boucaud:2011eh}.    

A significant part of the existing SDE analysis has focused on the study of  
various aspect of the aforementioned Green's functions at the 
level of pure gauge Yang-Mills theories, {\it i.e.}, without the inclusion 
of quarks~\cite{Aguilar:2008xm,Boucaud:2008ji,Dudal:2008sp}.
This tendency has been mainly motivated by the fact that the vast 
majority of lattice simulations work in the 
quenched limit, making no reference to effects 
stemming from dynamical quarks~\cite{Cucchieri:2007md,Bogolubsky:2007ud,Oliveira:2008uf}.  

The transition from pure $SU(3)$ Yang-Mills Green's functions to those of  real world QCD is, 
of course, highly nontrivial, and has been the focal point of relatively few
lattice investigations~\cite{Kamleh:2007ud,Bowman:2007du}. 
At the level of the SDEs, to the best of or knowledge, 
this issue has been studied in detail~\cite{Fischer:2003rp,Fischer:2005en}
only in the context 
of the so-called ``scaling solutions"~\cite{Fischer:2006ub}, 
but no analogous investigation has been carried out for the (IR finite) massive solutions~\cite{Cornwall:1981zr}, found both in the lattice simulations and in several of the analytic studies cited above.

The purpose  of the  present article is  to provide  a self-consistent
framework for addressing this latter  problem in the continuum, at the
level of  the corresponding  SDEs. In particular,  we will  present an
approximate  method for  ``unquenching'' the  (IR  finite) gluon
propagator  (in  the Landau  gauge),  computing nonperturbatively  the
effects induced by a small number of light quark families.

The method we present consists of two basic steps:
\n{i} computing the fully-dressed quark-loop diagram [see graph $(a_{11})$ in Fig.~\ref{glSDE}], using as input the nonperturbative quark propagators 
obtained from the solution of the gap equation, together with an Ansatz for  
the fully-dressed quark-gluon vertex that preserves gauge-invariance~\cite{Aguilar:2010cn}; and 
\n{ii} adding the result computed in \n{i} to the quenched gluon propagator obtained in 
the large-volume lattice simulations mentioned above~\cite{Bogolubsky:2007ud}.
The key assumption of the method employed is that    
the effects of a small number of quark families to the gluon propagator 
may be considered as a ``perturbation'' to the quenched case, and that   
the diagram $(a_{11})$ constitutes the leading correction. 
The subleading corrections stem   
from the (originally) pure Yang-Mills diagrams [graphs $(a_{1})$--$(a_{10})$ in Fig.~\ref{glSDE}], 
which now get modified 
from the quark loops nested inside them (see Fig.~\ref{unquenching}); 
their proper inclusion, however, lies beyond our present calculation powers.
So, our operating assumption is that these latter effects are small compared 
to those originating from graph $(a_{11})$, and will be neglected at this level of approximation.
It is interesting to note that in the context of the ``scaling'' solutions
this latter assumption appears to be indeed reasonable~\cite{Fischer:2003rp,Fischer:2005en}.

This assumption becomes relevant when implementing point \n{ii}, where the 
contributions from graphs $(a_{1})$--$(a_{10})$  will be taken to be exactly the same as those of the 
quenched case {\it even when dynamical quarks are present}, 
thus identifying  with the quenched lattice propagator everything except graph $(a_{11})$.  
Of course, as is typical in the SDE studies, the validity 
of this central assumption 
may be tested only a-posteriori, 
either by means of additional, more complicated computations, or, more pragmatically, 
through the levels of agreement achieved with available lattice results. As we will see 
in the main body of the article [Section {\ref{numres}-D], the general features emerging from our calculations are 
consistent with the lattice results of~\cite{Kamleh:2007ud,Bowman:2007du}.

The general framework we will adopt is based on the synthesis of the 
the pinch technique (PT)~\cite{Cornwall:1981zr,
Cornwall:1989gv,Binosi:2002ft,Binosi:2003rr,Binosi:2009qm}
with the background field method (BFM)~\cite{Abbott:1980hw}, 
known in the literature as the PT-BFM scheme~\cite{Aguilar:2006gr,Binosi:2007pi,Binosi:2008qk}.
As has been explained in detail in various works, the PT-BFM 
Green's functions satisfy Abelian-like Ward identities (WIs), 
instead of the typical Slavnov-Taylor identities (STIs), valid within the 
linear covariant ($R_{\xi}$) gauges~\cite{Abbott:1980hw,Binosi:2009qm}. The main consequence of this property is that the resulting SDE for the gluon self-energy may be suitably truncated, 
without compromising the transversality of the answer~\cite{Aguilar:2006gr,Binosi:2007pi,Binosi:2008qk}.

For the case at hand, the new ingredient is the nonperturbative quark loop,
which is transverse in the PT-BFM scheme as well as in the $R_{\xi}$ gauges; 
thus, at first sight, it would seem that there is no real advantage in using the former scheme.
However, the important issue at this point is the exact way how this transversality is 
realized in both cases. In particular, 
the fact that the fully dressed quark-gluon vertex 
of the PT-BFM (denoted by $\widehat{\Gamma}_{\mu}$)
satisfies a QED-like WI provides a definite advantage over 
the corresponding conventional vertex (denoted by ${\Gamma_{\mu}}$), 
which satisfies the STI 
that involves the quark-ghost scattering kernel~\cite{Marciano:1977su}, a relatively unexplored quantity
(see \2eqs{WI}{STI}, respectively)~\cite{Aguilar:2010cn}.
The reason why this constitutes an advantage has to do with the fact that, according to the common practice, 
one must eventually introduce a suitable nonperturbative Ansatz 
for the full quark-gluon vertex, such that the corresponding WIs (or STIs)
are automatically satisfied.
The fact that the PT-BFM vertex satisfies a WI instead of an STI simplifies 
the problem considerably, because it allows one to employ the time-honored Abelian 
Ans\"atze existing in the literature~\cite{Ball:1980ay,Curtis:1990zs}.

The necessary transition from the PT-BFM to the conventional gluon propagator, which is 
the one simulated on the lattice, is accomplished by means of a special Green's 
function, usually denoted by $G$ in the literature~\cite{Binosi:2007pi,
Binosi:2008qk}.
In the Landau gauge, $G$ is known to coincide with the ``Kugo-Ojima'' function, 
and to be related to the ghost dressing function by means of a powerful identity enforced by the underlying Becchi-Rouet-Stora-Tyutin (BRST) symmetry~\cite{Grassi:2004yq,Aguilar:2009pp}.
Thus, the use of the PT-BFM scheme eliminates the need to   
refer to quantities such as the quark-ghost kernel, at the 
very modest price of introducing the aforementioned function, which, due to its 
STI, can be accurately reconstructed from large-volume lattice data on the 
ghost dressing functions~\cite{Aguilar:2009pp,Aguilar:2009nf}, or possibly through 
direct lattice simulations of the Kugo-Ojima function~\cite{Sternbeck:2006rd}.  

The main results of our study may be summarized as follows.
The basic effect of the quark loop(s) (one or two families with a constituent mass of the order of 300 MeV)  
is to suppress considerably the gluon propagator in the IR and intermediate momenta regions, 
while the ultraviolet (UV) tails increase, exactly as expected from the standard renormalization group analysis. 
The final saturation point of the unquenched propagator cannot be reliably calculated at present; the 
apparent tendency is that 
the inclusion of light quarks makes the gluon propagator saturate at a lower point,  
which can be translated into having a larger gluon mass. 
We emphasize that the way the quark loops affect 
the value of the gluon mass is indirect: 
the contribution obtained from  graph $(a_{11})$  
vanishes at $q^2=0$, so it does not change the gluon mass equation formally~\cite{Aguilar:2011ux}; however,
it does change its solutions, because of the modification that it induces in the intermediate region of the 
gluon propagator (which enters in the gluon mass equation). 
A reliable estimate of this gluon mass difference cannot be obtained without resorting to the full gluon mass equation, 
whose derivation  is currently underway. For the purposes of the present work, the IR ``saturation point'' of the 
unquenched propagator will be estimated only approximately, 
through a processes of  ``extrapolation'' of the intermediate momenta region towards the deep IR.
The dependence of the results on the renormalization point $\mu$ is also studied in detail, and appears to be 
consistent with expectations based on general considerations. 

In addition, we present a direct comparison between unquenched gluon ``dressing functions'', namely 
the one obtained using the method described above with that found on the lattice~\cite{Kamleh:2007ud,Bowman:2007du}.
Note that, due to its very definition, the dressing function is rather insensitive to the exact value
of the final saturation point, moderating to some extent the effect of the aforementioned uncertainty. 
The resulting comparison with the lattice data is rather favorable, as may be seen in  Fig.~\ref{fig:lattice}; 
in the momentum region of maximum discrepancy the two curves differ by about 10\%, being significantly closer everywhere else.

Finally, it is quite interesting to mention that the use of the perturbative result for the quark loop [see Appendix] 
gives rise to an effect that is numerically very close
to that obtained through the more sophisticated field-theoretic treatment described above, as can be appreciated  
on the left panel of Fig.~\ref{fig:1loop}.

The article is organized as follows.
In section \ref{quarkloops} we give a detailed presentation of the basic methodology, main ingredients, and 
central assumptions of the procedure employed for adding quark loops to the  gluon propagator. 
In section \ref{nonpertloops} we elaborate on the way how the quark loop is computed nonperturbatively.
The main points of this section include \n{i} the particular form(s) of the full quark-gluon vertex employed, 
\n{ii} the actual computation of the loop and its behavior at  $q^2=0$, 
\n{iii} the (subtractive) renormalization procedure, and \n{iv} the transition to the Euclidean space. 
Section {\ref{numres} contains the main results of the present work. After introducing the
lattice ingredients used as input in our basic formulas, we present the unquenched gluon propagator for $SU(3)$, 
together with the corresponding dressing function, for a small number of light quark families. 
Further relevant points, such as the dependence of the results on the renormalization point, 
as well as the effect of ``decoupling'' of the heavy quarks are also addressed. 
In addition, a comparison of the resulting gluon dressing function with available lattice data~\cite{Kamleh:2007ud,Bowman:2007du} 
is presented. Our main conclusions and further open questions are summarized in section \ref{conc}. 
Finally, some useful formulas related to the perturbative (one-loop) calculation of the 
quark loop are summarized in an Appendix.

\section{\label{quarkloops}Adding quark loops to the gluon propagator}

%%%%%%%%%%%%%%%%%%%%%%%%%%%%%%%%%%%%%%%%%%%%%%%%%
%         figure 1 - SDE 
%%%%%%%%%%%%%%%%%%%%%%%%%%%%%%%%%%%%%
\begin{figure}[!t] 
\includegraphics[scale=.45]{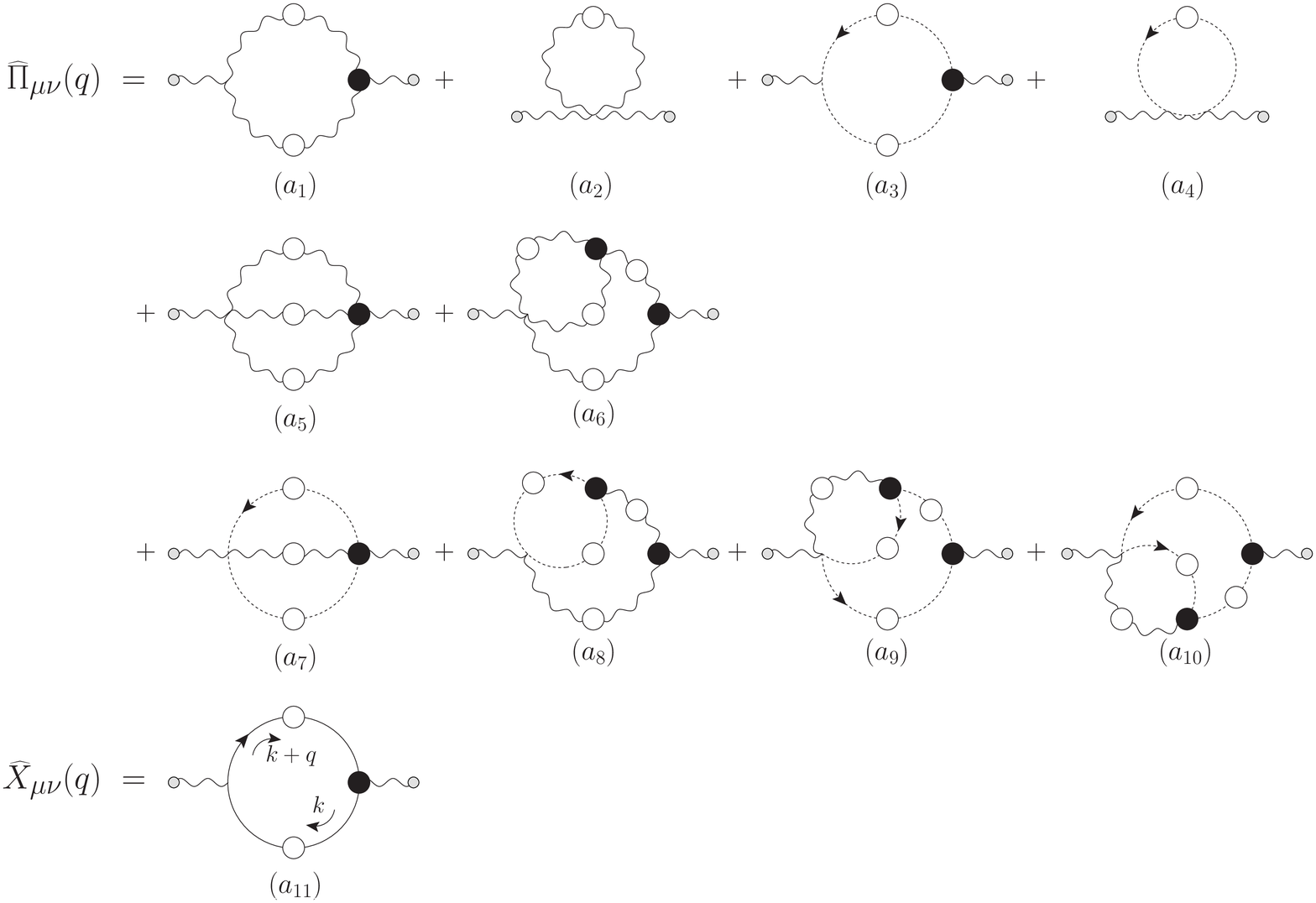}
\caption{\label{glSDE} The full PT-BFM gluon self-energy. White (respectively black) blobs represents connected (respectively 1-particle irreducible) Green's functions; the small gray circles on the external legs indicate background gluons.}
\end{figure}    
%%%%%%%%%%%%%%%%%%%%%%%%%%%%%%%%%%%

To begin with, in the Landau gauge the gluon propagator (quenched or unquenched) assumes the form
\be
\Delta_{\mu\nu}(q)=-i\Delta(q^2)P_{\mu\nu}(q); \qquad
P_{\mu\nu}(q)=g_{\mu\nu}-\frac{q_\mu q_\nu}{q^2}.
\ee
Let us now denote by $\Delta_{\s{Q}}(q^2)$
the full gluon propagator 
in the presence of quark loops, while the 
corresponding quenched propagator, \ie 
the full gluon propagator in the absence of quark loops, 
will be denoted simply by $\Delta(q^2)$.

%%%%%%%%%%%%%%%%%%%%%%%%%%%%%%%%%%%%%%%%%%%%%%%%%%
%              Figure 2 - H 
%%%%%%%%%%%%%%%%%%%%%%%%%%%%%%%%%%%%%%%%%%
\begin{figure}[!t]
\includegraphics[scale=.55]{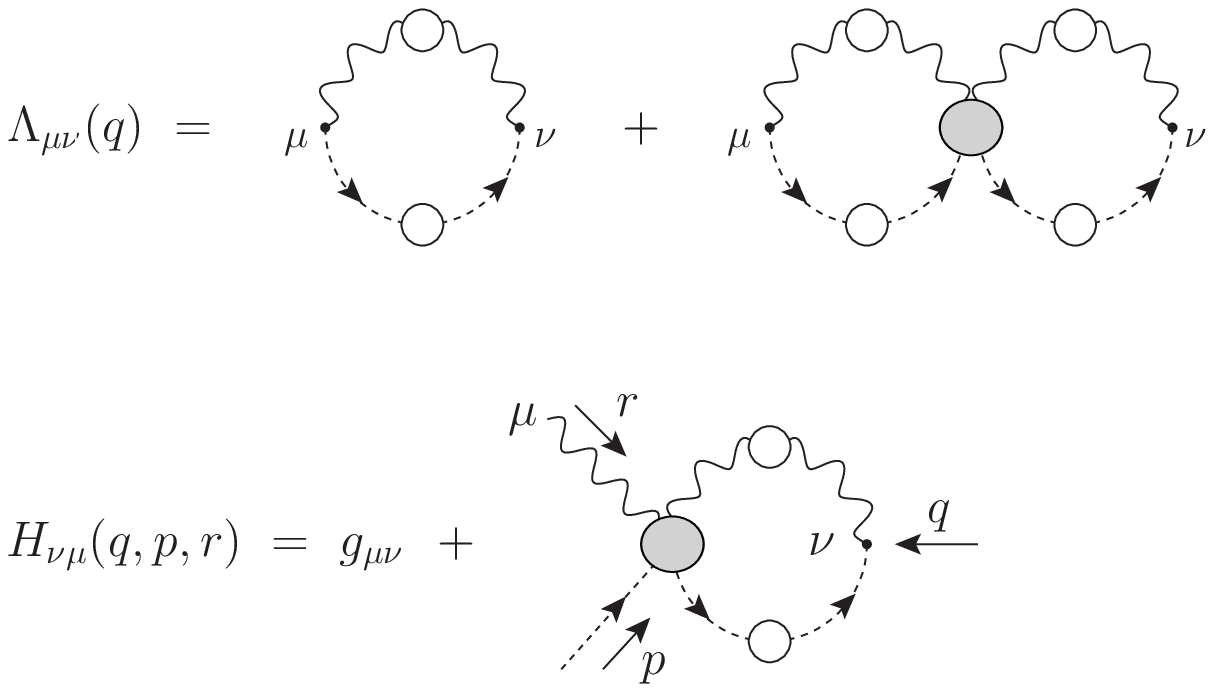}
\caption{\label{Lambda-H}Definitions and conventions of the auxiliary functions $\Lambda$ and $H$.
The color and gauge coupling dependence for the field combination shown, $c^a(p)A_\mu^b(r)A_\nu^{*c}(q)$, is $gf^{acb}$.  Gray blobs denote  one-particle irreducible (with respect to vertical cuts) Schwinger-Dyson kernels.}
\end{figure}
%%%%%%%%%%%%%%%%%%%%%%%%%%%%%%%%%%%%%%%%%%% 

In the PT-BFM scheme, $\Delta(q^2)$ satisfies the following 
SDE~\cite{Binosi:2007pi,Aguilar:2008xm,Binosi:2009qm},
\be
\Delta^{-1}(q^2)P^{\mu\nu}(q)= \frac{q^2P^{\mu\nu}(q)+i \widehat\Pi^{\mu\nu}(q)}
{\left[1+G(q^2)\right]^2}. 
\label{Dr}
\ee
where 
\be
\widehat\Pi^{\mu\nu}(q)= \sum_{i=1}^{10} (a_{i})^{\mu\nu},
\ee
and the relevant fully dressed diagrams $(a_i)$ are shown in \fig{glSDE}.
All these diagrams contain only fields appearing in 
the pure gauge Yang-Mills Lagrangian, namely gluons and ghosts. 
The function $G$ appearing in~\noeq{Dr} is particular to the PT-BFM formalism~\cite{Binosi:2007pi,
Binosi:2008qk}; specifically, 
it is the form factor associated with the metric tensor $g_{\mu\nu}$
in the Lorentz decomposition of the auxiliary two-point function $\Lambda_{\mu\nu}$, given by~\cite{Binosi:2009qm}
\bea
\Lambda_{\mu\nu}(q)&=&-ig^2C_A\int_k\!\Delta_\mu^\sigma(k)D(q-k)H_{\nu\sigma}(-q,q-k,k)\nonumber\\
&=&g_{\mu\nu}G(q^2)+\frac{q_\mu q_\nu}{q^2}L(q^2).
\label{defG}
\eea
In the formula above, $C_A$ is the Casimir eigenvalue in the adjoint representation 
[$C_A=N$ for $SU(N)$], and
the $d$-dimensional integral (in dimensional regularization) is defined according to
\be
\int_{k}\equiv\frac{\mu^{\epsilon}}{(2\pi)^{d}}\!\int\! \diff{d} k, 
\label{dqd}
\ee
with $d= 4-\epsilon$ and $\mu$ the 't Hooft mass. 
The function $\Lambda_{\mu\nu}(q)$, together with the auxiliary function $H_{\mu\nu}(q,p,r)$, 
are diagrammatically represented  in Fig.~\ref{Lambda-H}.

Notice that  $H_{\mu\nu}$ is  related to the (conventional) gluon-ghost $\Gamma^c_{\mu}$ vertex by the identity
\be
p^\nu H_{\nu\mu}(p,r,q)+\Gamma^c_{\mu}(r,q,p)=0,
\ee
and that, in the (background) Landau gauge,  
 the following all order relation holds~\cite{Grassi:2004yq,Aguilar:2009pp}
\be
F^{-1}(q^2)=1+G(q^2)+L(q^2).
\label{funrel}
\ee 

The unquenched propagator in the presence of a single quark loop 
will satisfy an appropriately modified version of (\ref{Dr}), namely 
\be
\Delta_{\s{Q}}^{-1}(q^2) P^{\mu\nu}(q)= 
\frac{q^2P^{\mu\nu}(q)+i \widehat\Pi^{\mu\nu}_{\s{Q}}(q)  + 
i\quark^{\mu\nu}(q)}{\left[1+G_{\s{Q}}(q^2)\right]^2}. 
\label{Drnf}
\ee

%%%%%%%%%%%%%%%%%%%%%%%%%%%%%%%%%%%
%             Fig.3 propagation nonlinear effect
%%%%%%%%%%%%%%%%%%%%%%%%%%%%%%%%
\begin{figure}[!t] 
\includegraphics[scale=.73]{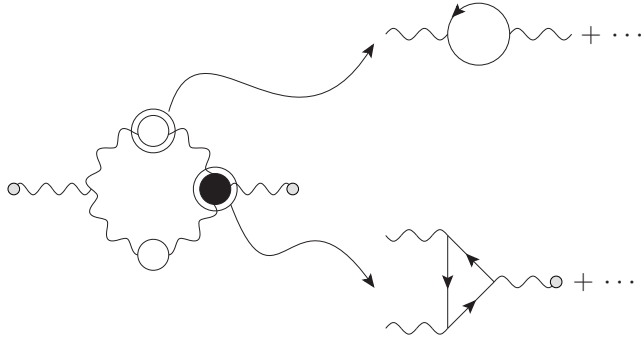}
\caption{\label{unquenching}The nonlinear propagation of the effect of unquenching the gluon propagator through the addition of dynamical fermions, shown here for the one-loop dressed gluon diagram $(a_1)$. Both the internal gluon propagator and the three-gluon vertex gets modified (shown here by  two representative graphs only); similar modifications occur for all other diagrams.}
\end{figure}
%%%%%%%%%%%%%%%%%%%%%%%%%%%%%%%%%%%%%%%%%%%%%

The main difference between~\noeq{Dr} and~\noeq{Drnf}
is the explicit appearance of $\quark^{\mu\nu}(q)$ on the rhs, originating entirely
from diagram $(a_{11})$ (see again \fig{glSDE}). Notice, however, that there will be a nonlinear propagation of the 
changes induced due to $\quark^{\mu\nu}(q)$, which will also affect the original subset 
of purely Yang-Mills graphs, namely $(a_1)-(a_{10})$, given that 
now the various Green's functions appearing inside them 
will have been modified by $\quark^{\mu\nu}(q)$. For example, at the ``one-loop dressed'' level, diagram $(a_1)$  
receives quark-loop contributions, such as those shown in \fig{unquenching}, and the same happens with all 
other graphs belonging to the set $(a_1)-(a_{10})$. 
This complicated nonlinear effect is indicated by introducing 
the suffix ${Q}$ in the associated self-energy, 
$\widehat\Pi^{\mu\nu}_{\s{Q}}(q)$. 
The quantity $G(q)$ will be similarly affected by the inclusion of the quark loop, as indicated 
in (\ref{Drnf}) through the substitution  $G(q) \to G_{\s{Q}}(q)$.
In the case of including various quark loops, corresponding to different quark flavors, 
$Q_i$, the term $\quark^{\mu\nu}(q)$ in (\ref{Drnf})  
is replaced simply by the sum over all quark loops, \ie 
\be
\quark^{\mu\nu}(q) \to \sum_{i}\widehat{X}_{i}(q) \,.
\ee
The tensorial structure in \2eqs{Dr}{Drnf} may be easily eliminated, 
by appealing to the transversality properties of the quantities involved on the rhs, namely 
\be
q_{\mu} \widehat\Pi^{\mu\nu}(q)= 0;
\qquad
q_{\mu} \quark^{\mu\nu}(q) = 0;
\qquad
q_{\mu} \widehat\Pi^{\mu\nu}_{\s{Q}}(q) = 0.
\ee
Let us now define the scalar cofactors of these quantities as  
\be
\widehat\Pi^{\mu\nu}(q) =  P^{\mu\nu}(q) \widehat\Pi(q^2);
\qquad
\quark^{\mu\nu}(q) =  P^{\mu\nu}(q) \quark(q^2);
\qquad
\widehat\Pi^{\mu\nu}_{\s{Q}}(q) = P^{\mu\nu}(q) \widehat\Pi_{\s{Q}}(q^2).
\ee
Then,  
equations (\ref{Dr}) and (\ref{Drnf}) can be converted to their scalar versions, namely    
\be
\Delta^{-1}(q^2) = \frac{q^2 +i \widehat\Pi(q^2)} 
{\left[1+G(q^2)\right]^2} \,,
\label{Drsc}
\ee
and 
\be
\Delta_{\s{Q}}^{-1}(q^2) = 
\frac{q^2 + i \widehat\Pi_{\s{Q}}(q^2) 
+i \quark(q^2)}{\left[1+G_{\s{Q}}(q^2)\right]^2}. 
\label{Drnfsc}
\ee
Eq.~(\ref{Drnfsc}) can be then straightforwardly adjusted to include the case of various quark loops, simply by replacing 
$\quark(q)\to \sum_i \widehat{X}_{\s{Q_i}}(q)$. 

To be sure, 
the total effect of including quark loops 
cannot be exactly computed at the level of the SDE, because that would entail 
the full numerical treatment of the entire series, a task that is beyond our present powers.
The way we will proceed instead is the following. We will use the quenched propagator 
as our reference, and we will estimate the modifications introduced to it by the presence 
of the quark loop(s), under certain simplifying assumptions that we will now explain.

To that end, let us cast the quenched 
gluon propagator $\Delta (q^2)$ into the standard form employed in the recent 
literature~\cite{Aguilar:2009ke,Aguilar:2011ux, Aguilar:2011xe},
which incorporates the crucial feature of IR finiteness, 
implemented by the presence of a dynamically generated gluon mass; specifically, we set (in Minkowski space),  
\be
\Delta^{-1}(q^2)=q^2 J(q^2)-m^2(q^2).
\label{h1}
\ee
The first term on the rhs of (\ref{h1}) corresponds to   the ``kinetic term'', or ``wave function'' contribution, 
whereas the second is the  momentum-dependent mass 
(which is positive-definite in Euclidean space)~\cite{Aguilar:2009ke,Aguilar:2011ux,Aguilar:2011xe}}. 
As \mbox{$q^2\to 0$}, we have that  \mbox{$q^2 J_m(q^2) \to 0$}; on the other hand,  \mbox{$m^2(0) \neq 0$}, and as a result, 
the gluon propagator is IR finite, \mbox{$\Delta^{-1}(0)\neq 0$}. 
The exact determination  of the components $J(q^2)$ and $m^2(q^2)$ 
in terms of the quantities appearing on the rhs of (\ref{Dr}) and (\ref{Drnf}) is a complicated task,
leading eventually to a set of intricate coupled integral equations. This exercise has been carried out partially, within the one-loop truncated version of the SDE,  
considering only the corresponding subset of gluonic contributions [\ie diagrams $(a_1)$ and $(a_2)$]~\cite{Aguilar:2011ux}. 

In what follows we will operate under the reasonable assumption that the IR 
finiteness of the gluon propagator persists in the presence of a 
relatively small number 
of quark loops. In other words, 
we assume that the inclusion of two light quark flavors (up and down type quarks,
with constituent masses of about \mbox{$300$ MeV})
will affect but not completely destabilize the mechanism responsible for 
the generation of a dynamical gluon mass, and that their effect may be considered as a ``perturbation'' 
to the quenched case. In the realistic case of QCD, 
the inclusion of loops containing the remaining heavier quarks   
is expected to give rise to numerically suppressed contributions 
(compared to those coming from the 
light quark loops), consistent with the notion of decoupling; this 
expectation is in fact clearly confirmed 
in the results presented in Section~\ref{numres} (see in particular Fig.~\ref{fig:1loop}). 
Instead, the theoretical possibility of increasing the number of loops containing light flavors 
may lead to effects that cannot be longer considered as a ``perturbation'' of the quenched case: 
ten families of light quarks, for example, could alter severely the qualitative behavior of the 
theory, and as a result, the quenched propagator may have little to do with the unquenched one
(for a general discussion on how the IR and UV properties of Yang-Mills theories may be distorted, depending on the 
number of quark families, see, e.g.,~\cite{DelDebbio:2010zz,Cheng:2011qc}, and references therein). 

Thus, under the aforementioned assumptions, \1eq{h1} will be extended to the case of $\Delta_\s{Q}(q^2)$, namely  
\be
\Delta^{-1}_\s{Q}(q^2)=q^2 J_\s{Q}(q^2)-m^2_\s{Q}(q^2),
\label{h2}
\ee
where the suffix $Q$ in the dynamical gluon mass indicates the possible modifications to $m^2(q^2)$ 
induced by the quark loop(s), as alluded above. It is important to emphasize that $m^2(q^2)$  
will change, despite the fact that the main additional ingredient 
that distinguishes (\ref{Drsc}) and (\ref{Drnfsc}), namely  $\quark(q)$, 
does not contribute at $q^2=0$,  since $\quark(0) =0$ [see \1eq{X0}],
and therefore it does not affect {\it directly} the gluon mass equation~\cite{Aguilar:2011ux}; instead, the modification 
induced is {\it indirect}, due to the change in the overall shape of $\Delta(q^2)$ 
throughout the entire range of momenta. In order to gain a qualitative understanding of this last statement, 
let us consider the IR limit of the approximate gluon mass equation 
obtained in~\cite{Aguilar:2011ux}, where only the one-loop dressed graphs $(a_1)$ and $(a_2)$ are considered; in Euclidean space, 
\be
m^2(0) 
= -\frac{3C_A}{8\pi}\alpha_s F(0) \int_0^\infty\!\diff{} y\,m^2(y) [\dress^2(y)]^{\prime}  + \dots,
\label{m20}
\ee 
where \mbox{$\alpha_s=g^2/4\pi$}, the prime indicates differentiation with respect to $y=k^2$,   
and $\dress(y)$ is the ``dressing function'' of the gluon propagator, defines as 
\be
\dress(q^2) \equiv q^2 \Delta(q^2).
\label{dress1}
\ee 
Evidently, $\dress(0) =0$. Finally, 
the ellipses on the rhs of \1eq{m20} denote contributions from ``two-loop dressed'' diagrams that have yet to be worked out.

Now, in the presence of quark loops, \1eq{m20} maintains its functional form, 
since, as mentioned above, ${\quark}(0) =0$; however, the various quantities appearing on its rhs  
[most notably  $\dress(y)$] will be modified, therefore acquiring a suffix ``Q'' [{\it e.g.}, $\dress(y)\to\dress_\s{Q}(y) $]. 
As a consequence, the 
resulting solution gets modified, and we have $m^2(q^2)\to m^2_\s{Q}(q^2)$; in what follows we will denote by 
\be
\lambda^2 \equiv m^2_{\s{Q}}(0) - m^2(0),
\label{lambda}
\ee
the gluon mass difference at $q^2 =0$.

As already explained, a solid first-principle determination of $\lambda^2$  
is not possible at the moment, mainly due to the 
fact that the available gluon mass equation~\noeq{m20} is incomplete, since it has been derived from only one subset of the 
relevant graphs~\cite{Aguilar:2011ux}. Therefore, in the analysis presented we will restrict ourselves to extracting  
an approximate range for $\lambda^2$, through the extrapolation of the curves obtained from intermediate momenta
towards the deep IR.

In order to estimate the effect of the quark loop(s) on the gluon propagator, we will 
assume that the main bulk of the correction to the ``kinetic'' part, $q^2 J_{\s{Q}}(q^2)$, 
is due to the direct presence of the extra diagram $(a_{11})$. 
Instead, the nonlinear 
effect due to the fact that the graphs $(a_1)-(a_{10})$ 
develop an indirect quark dependence, \ie $\widehat\Pi(q^2) \to \widehat\Pi_{\s{Q}}(q^2)$, 
is predominantly responsible for 
the change in the gluon mass, as captured in (\ref{lambda}), inducing 
minor changes to the kinetic part $q^2 J_{\s{Q}}(q^2)$. 
Finally, we will approximate the function $G_{\s{Q}}(q^2)$ appearing in the denominator of  
\1eq{Drnfsc} by the quenched expression, \ie \mbox{$G_{\s{Q}}(q^2)\to G(q^2)$}; as can be seen from its  
defining equation \1eq{defG} and Fig.~\ref{Lambda-H}, quark-loops enter only as ``higher order'' effects,  
according to our general philosophy, and their effect should be small. 

Thus, within this approximation scheme, the quantity $J_{\s{Q}}(q^2)$ will be given by  
\be
q^2 J_{\s{Q}}(q^2) = q^2 J(q^2) + \frac{i \,\quark(q^2)}{\left[1+G(q^2)\right]^2}.
\label{JQ}
\ee
If we now combine Eqs.~(\ref{h1}), (\ref{h2}), (\ref{m20}) and (\ref{lambda}),  
it is easy to arrive at the result (Minkowski space)
\be
\Delta_{\s{Q}}(q^2) = \frac{\Delta(q^2)}
{1 + \left\{ i \,\quark(q^2) \left[1+G(q^2)\right]^{-2}- \lambda^2 \right\}\Delta(q^2)}.
\label{mastform}
\ee
In what follows we will identify the quenched propagator $\Delta(q^2)$ appearing on the rhs of (\ref{mastform})
with the one obtained from the large volume lattice simulations~\cite{Bogolubsky:2007ud}, to be denoted by $\Delta_\s{L}(q^2)$.
So, effectively one assumes that $\Delta_\s{L}(q^2)$ is a solution of the full SDE equation, with no quarks, given 
in (\ref{Drsc}); thus, when using (\ref{mastform}) 
we will be carrying out the replacement $\Delta(q^2)\to \Delta_\s{L}(q^2)$. 

\section{\label{nonpertloops}Nonperturbative Quark loop in the PT-BFM scheme}

In this section we present the actual nonperturbative calculation of the quark-loop diagram $(a_{11})$, finally expressing the answer exclusively in terms of  
the functions $A(p)$ and $B(p)$, appearing in the Dirac decomposition of the full 
quark propagator [see (\ref{qprop})]. The calculation relies on the use of suitable Ans\"atze for 
the fully dressed quark-gluon vertex appearing in $(a_{11})$, presented and discussed in the corresponding subsection.
The Euclidean version of the (renormalized) 
master formula that we use in the next section in order to estimate the effect of the quark loop on the 
gluon propagator is given   in Eq.~(\ref{mastformeuc}).

\subsection{The quark-gluon vertex}

%%%%%%%%%%%%%%%%%%%%%%%%%%%%%%%%%%%%%%%%%%
%          Fig.4  - vertex 
%%%%%%%%%%%%%%%%%%%%%%%%%%%%%%%%%%%%%%%
\begin{figure}[!t]
\includegraphics[scale=.70]{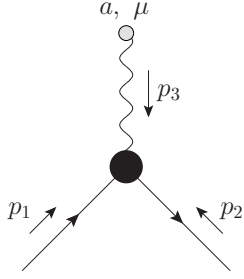} 
\caption{\label{quark-vertex}The full PT-BFM quark-gluon vertex $\widehat\bqq^a_\mu$.  Note 
that $p_1 \leftrightarrow p_2$ with respect to the conventions used in~\cite{Davydychev:2000rt,Aguilar:2010cn}.}
\end{figure}
%%%%%%%%%%%%%%%%%%%%%%%%%%%%%%%

The quantity responsible for the difference between the quark loop
in the conventional covariant gauges and the PT-BFM scheme  is the fully-dressed quark-gluon vertex.
Specifically, let us denote the fully dressed PT-BFM quark-gluon vertex by $\widehat\bqq^a_\mu$, 
and factor out the color structure, according to 
\be
\widehat\bqq^a_\mu (p_1,p_2,p_3) = g\,t^a\widehat\G_\mu(p_1,p_2,p_3),
\ee
where all momenta $p_i$ entering (see Fig.~\ref{quark-vertex}); 
at tree-level, $\widehat\G^{(0)}_\mu=\gamma_\mu$.
In the equation above, $t^a$ represents the $N^2-1$ hermitian and traceless generators of the $SU(N)$ gauge group, satisfying the algebra
\be
[t^a,t^b]=if^{abc}t^c,
\ee
with $f^{abc}$ the totally antisymmetric group structure constants. 
In the $SU(3)$ case, with 
the quarks in the fundamental representation, $t^a=\lambda^a/2$, 
where $\lambda^a$ are the Gell-Mann matrices.

In the conventional formulation within the linear covariant gauges, 
the quark-gluon vertex, to be denoted by $\G_\mu(p_1,p_2,p_3)$, satisfies the well-known 
STI~\cite{Marciano:1977su}
\be
ip_3^{\mu}\Gamma_{\mu}(p_1,p_2,p_3) = 
F(p_3)[S^{-1}(p_1) H(p_2,p_1,p_3) - {\overline H}(p_1,p_2,p_3) S^{-1}(-p_2)],
\label{STI}
\ee
where $S^{-1}(p)$ is the inverse of the full quark propagator, 
$H(p_2,p_1,p_3)$ is the quark-ghost scattering kernel diagrammatically defined in Fig.~\ref{qg_H},
and  ${\overline H}(p_1,p_2,p_3)$ its ``conjugate''.

In contrast, 
in the PT-BFM scheme, the vertex $\widehat\G_\mu$ satisfies the QED-like WI~\cite{Abbott:1980hw,Binosi:2009qm} 
\be
ip_3^\mu\widehat\G_\mu(p_1,p_2,p_3)=S^{-1}(p_1)-S^{-1}(-p_2), 
\label{WI}
\ee
with no reference whatsoever to the ghost sector.  Then, 
the most general Ansatz for the longitudinal part of $\widehat\G_\mu$ that satisfies (\ref{WI}) 
is given by~\cite{Ball:1980ay}
\be
\widehat\G_\mu(p_1,p_2,p_3)=L_1(p_1,p_2)\gamma_\mu+L_2(p_1,p_2)
\left(p_1\!\!\!\!\!\slash-p_2\!\!\!\!\!\slash\,\,\right)\left(p_1-p_2\right)_\mu+L_3(p_1,p_2)\left(p_1-p_2\right)_\mu.
\label{full-vertex}
\ee
The form factors $L_i$ appearing in the expression above are given by  
\be
L_1(p_1,p_2)=\frac{A(p_1)+A(p_2)}{2}; \quad
L_2(p_1,p_2)=\frac{A(p_1)-A(p_2)}{2\left(p_1^2-p_2^2\right)};\quad
L_3(p_1,p_2)=-\frac{B(p_1)-B(p_2)}{p_1^2-p_2^2}.
\label{theLs}
\ee
where the functions $A(p)$ and $B(p)$ are defined as 
\be
S^{-1}(p)=-i\left[A(p)\pslash-B(p)\right]=-iA(p)\left[\pslash-\qm(p)\right] \,,
\label{qprop}
\ee
and the ratio \mbox{$\qm(p)=B(p)/A(p)$} is the dynamical quark mass function.
For latter convenience, we will denote the dynamical quark mass function at \mbox{$p^2=0$} by \mbox{$\qm(0) \equiv M$}. 
Therefore, at tree level ($A=1$, $B=M$) and one has $L_1=1$ and $L_2=L_3=0$.
The resulting vertex reads 
\bea 
\widehat\G_\mu(p_1,p_2,p_3)&=&\frac{A(p_1)+A(p_2)}{2}\gamma^{\mu} 
\nonumber \\
&+&\frac{(p_1-p_2)^{\mu}}{p_1^2-p_2^2}\left\{\left[A(p_1)-A(p_2)\right] 
\frac{\sla{p_1}-\sla{p_2}}{2}
+\left[B(p_1)-B(p_2)\right] 
\right\} \,.
\label{bcvertex}
\eea
and is known in the literature as the Ball-Chiu (BC) vertex~\cite{Ball:1980ay}.

%%%%%%%%%%%%%%%%%%%%%%%%%%%%%%%%%%%%%%%%%%
%          Fig.4b  - quark-ghost scattering 
%%%%%%%%%%%%%%%%%%%%%%%%%%%%%%%%%%%%%%%
\begin{figure}[!t]
\includegraphics[scale=.55]{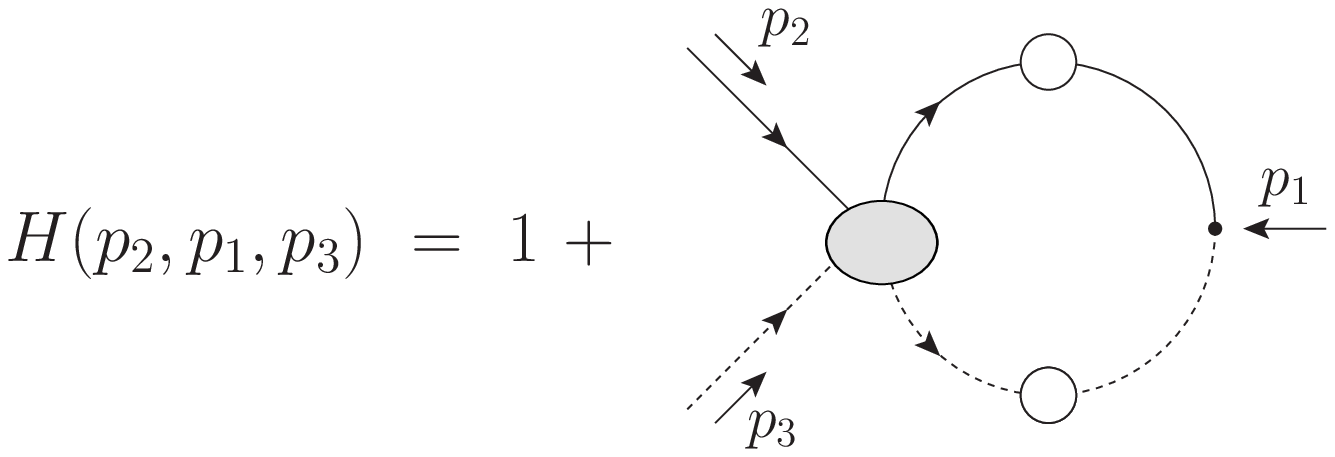} 
\caption{\label{qg_H} Diagrammatic representation of the quark-ghost scattering kernel $H(p_1,p_2,p_3)$.}
\end{figure}
%%%%%%%%%%%%%%%%%%%%%%%%%%%%%%%

We emphasize that in the context of the PT-BFM the {\it longitudinal} part of the above vertex is complete, as far as the 
WI it satisfies is concerned. 
Indeed, the expression in~(\ref{bcvertex}) satisfies the {\it exact} WI that $\widehat\G_\mu$ is
supposed to obey, namely (\ref{WI}). 
As is well-known, the BC vertex has been employed extensively in the 
literature (especially in studies of chiral symmetry breaking)~\cite{Roberts:1994dr}
 as an approximate (denominated ``abelianized'') version of the conventional $\Gamma_{\mu}$ 
defined in the covariant gauges. Indeed, the fully dressed quark-gluon vertex entering into the quark gap equation 
is $\Gamma_{\mu}$ and not $\widehat\G_{\mu}$, for the simple reason that the corresponding 
gluon is quantum and not background; indeed, the gluon in the quark gap equation is internal 
({\it i.e.}, it is irrigated by the virtual momenta), in contrast to the gluon of the quark loop, 
which is external (carries physical momentum). Therefore, use of  
the expression given in (\ref{bcvertex}) into the quark gap equation 
constitutes only an approximation,  
since it fails to satisfy the full STI (\ref{STI}) that $\Gamma_{\mu}$ should obey, unless the 
corresponding ghost sector is turned off. 

Note that the BC vertex  
has been generalized accordingly in~\cite{Aguilar:2010cn}, in order to fulfill the exact STI~(\ref{STI}), 
thus justifying its use inside the quark gap equation.  
The corresponding $L_i$ are considerably  more complicated than those given in (\ref{theLs}), 
involving the ghost dressing function $F$ and the various form factors of 
the quark-ghost kernel $H(p_1,p_2,p_3)$~\cite{Aguilar:2010cn}.
In fact, an additional fourth form factor, $L_4$, makes its appearance in the 
Lorentz expansion corresponding to (\ref{full-vertex}), multiplying 
$\sigma_{\mu\nu}= i/2[\gamma_{\mu},\gamma_{\nu}]$;
it is then easy to verify that this latter, genuinely non-Abelian vertex of~\cite{Aguilar:2010cn}
reduces to that of (\ref{bcvertex})  
in the limit of a trivial ghost sector,  {\it i.e.}, by setting $F(p) =1$ and $H =1$.

Finally, let us comment on an alternative form of  the quark-gluon vertex $\widehat\G_\mu(p_1,p_2,p_3)$, 
known in the literature as the 
Curtis and Pennington (CP) vertex~\cite{Curtis:1990zs}, to be denoted by $\widehat\Gamma_{\mu}^{\s{\rm CP}}$. 
This latter vertex satisfies also the WI of (\ref{WI}), and differs from 
the vertex of (\ref{bcvertex}) by a transverse (automatically conserved) contribution, which 
improves its properties under multiplicative renormalizability.
Specifically,  
\be
\widehat\Gamma_{\mu}^{\s{\rm CP}}(p_1,p_2,p_3) = \widehat\Gamma_{\mu} (p_1,p_2,p_3) 
+  \left[\gamma_{\mu}(p_2^2-p_1^2) + (p_2-p_1)_{\mu} \sla{p_3}\right]\widehat{\Gamma}_{\!\s{\rm T}}(p_1,p_2,p_3),
\label{full_cp1}
\ee
where 
\be
\widehat{\Gamma}_{\!\s{\rm T}}(p_1,p_2,p_3) = 
\frac{\left[A(p_2)-A(p_1)\right](p_1^2+p_2^2)}{2\bigg\{(p_2^2-p_1^2)^2 + 
\left[ \qm^2(p_2) + \qm^2(p_1)\right]^2\bigg\} }\,.
\label{cps}
\ee

In the analysis that follows we will use both the BC and the CP vertices, and compare the 
difference they induce to the various quantities of interest.

\subsection{The quark loop}
Let us now turn to the quark-loop diagram $(a_{11})$ of the PT-BFM scheme.
Factoring out the trivial color structure $\delta^{ab}$, we obtain  
\be
\quark^{\mu\nu}(q^2)=-g^2\,d_f\!\kint\mathrm{Tr}\left[
\gamma^\mu S(k)\widehat\Gamma^\nu(k+q,-k,-q)S(k+q)\right] \,,
\label{qse}
\ee
where $d_f$ is the Dynkin index of the fundamental representation [$d_f=1/2$ for $SU(3)$].

Since by virtue of the WI~(\ref{WI}) the quark loop $\quark^{\mu\nu}(q^2)$ 
is transverse\footnote{Note that the corresponding quark-loop in the covariant gauges, 
\ie with $\widehat\Gamma^\nu \to \Gamma^\nu$ is also transverse, by virtue of the STI~(\ref{STI}).}, 
\be
q_{\mu}\quark^{\mu\nu}(q^2) =0, 
\ee
we have that $\quark^{\mu\nu}(q^2)=\quark(q^2)P^{\mu\nu}(q)$; then, 
contracting with $g_{\mu\nu}$, and setting $d_f=1/2$,  we obtain 
\be
\quark(q^2)=-\frac{g^2}{2(d-1)}\!\kint\mathrm{Tr}\left[
\gamma^\mu S(k)\widehat\Gamma_\mu(k+q,-k,-q)S(k+q)\right].
\label{quark-loop} 
\ee

After inserting the full vertex~(\ref{full-vertex}) into~(\ref{quark-loop}) 
and taking the trace, we find one term for each of the form factors $L_i$. Specifically, we have that 
\be
\quark(q^2)=-\frac{2 g^2}{d-1}\kint\,\frac{1}{A_a  A_b (k^2-\qm_a^2)[(k+q)^2-\qm_b^2]}\sum_{i=1}^3 T_i(k,k+q)\,,
\label{quark-loop-full}
\ee
where the subindex ``$a$'' (respectively, ``$b$'') indicates 
that the corresponding function is evaluated at momentum $k$ (respectively, $k+q$), with
\bea
T_1(k,k+q)&=& L_1\left\{(2-d)(k^2+k\cdot q)+ d\qm_a\qm_b\right\},\nonumber \\
T_2(k,k+q)&=& L_2\bigg\{2\left[k\cdot(2k+q)\right]\left[(k+q)\cdot(2k+q)\right]-k\cdot(k+q)(2k+q)^2\nonumber \\
&+&(2k+q)^2\qm_a\qm_b\bigg\},\nonumber \\
T_3(k,k+q)&=& L_3 \bigg\{\qm_b \,[(2k+q)\cdot k] +\,\qm_a \,[(2k+q)\cdot (k+q)]\bigg\}.
\label{T123}
\eea

Before studying in detail each term, let us consider $\quark(q^2)$ in the limit $q\to0$. 
Using the expressions given in \1eq{theLs}, 
and dropping the subindices (all quantities being evaluated at $k$ now), one finds
\be
\quark(0)=-\frac{2g^2}{d-1}\kint\,\frac1{A^2(k^2-\qm^2)^2}\Bigg\{A\left[(2-d)k^2+d\qm^2\right]+2A'k^2(k^2+\qm^2)-4k^2B'\qm\Bigg\}.
\label{quark-loop-full-0}
\ee  
The important point to recognize now is 
that the integral on the rhs of~\noeq{quark-loop-full-0}
vanishes by virtue of an identity 
valid in dimensional regularization. This identity, 
referred to as the ``seagull identity'' in the recent literature~\cite{Aguilar:2009ke} 
constitutes the generalization of the simple identity (\ref{id1})
employed in the Appendix for the one-loop perturbative result.
     
Specifically, the seagull identity reads 
\be
\int_k\! k^2 f'(k^2)+\frac{d}2\int_k\! f(k^2)=0,
\label{seagull}
\ee 
where the ``prime'' denotes differentiation with respect 
to $k^2$, {\it i.e.}, $f'(k^2)\equiv \frac{\mathrm{d}f(k^2)}{\mathrm{d} k^2}$. 
%The identity of (\ref{id1}) corresponds to the choice $f(k)= (k^2-m^2)^{-1}$.
Interestingly enough, using inside \1eq{seagull} the function
\be
f(k^2)= [A(k^2) (k^2-\qm^2(k^2))]^{-1},
\ee 
namely the all order generalization of the one-loop \mbox{$(k^2-M^2)^{-1}$} employed in \1eq{id1}, 
one obtains precisely the integral on the rhs of (\ref{quark-loop-full-0}); 
therefore, 
\be
\quark(0)= 0 \,,
\label{X0}
\ee  
as announced.

Let us now compute the quark self-energy $\quarkCP(q^2)$
obtained by substituting into (\ref{qse})
the vertex $\widehat\Gamma^{\mu}_{\rm CP}$, given in  (\ref{full_cp1})-(\ref{cps}).
The answer will be expressed as a deviation from $\quark(q^2)$, namely
\be
\quark^{\s \rm CP}(q^2) = \quark(q^2) + \delta \quark(q^2) \,,
\label{delta_cp}
\ee 
where 
\be
\delta \quark(q^2) =  2 g^2 \kint\,\frac{[\qm_a\qm_b -(k^2+k\cdot q)][(k+q)^2-k^2]}
{A_a  A_b (k^2-\qm_a^2)[(k+q)^2-\qm_b^2]} \,\widehat{\Gamma}_{\s{\!\rm T}}(k+q,-k,-q) \,,
\label{quarkCP}
\ee 
It is easy to verify that the integral on the rhs of (\ref{quarkCP}) vanishes in the limit 
 $q\to 0$, because, as can be seen directly from  (\ref{cps}), $\widehat{\Gamma}_{\s{\!\rm T}}(p_1,-p_1,0)=0$. 
Therefore, the property of (\ref{quark-loop-full-0}) persists, namely $\quark^{\s{\rm{CP}}}(0) =0$.  

\subsection{Renormalization}
Clearly $\quark(q^2)$ (and $\quark^{\s{\rm{CP}}}(q^2)$) must be renormalized within  
the momentum subtraction (MOM) scheme. This choice is dictated by the fact that 
our final results will be expressed as deviations from the 
quenched gluon propagator obtained from the lattice, 
where the latter scheme has been employed. 
The renormalized expression for $\quark(q^2)$ in the MOM scheme is given by 
\be
\quarkren(q^2) = \quark(q^2) - \frac{q^2}{\mu^2} \quark(\mu^2).
\label{Xren}
\ee

As far as the propagator of (\ref{mastform}) is concerned, 
its renormalization will proceed as follows. 
First of all, as happens almost exclusively 
at the level of SDEs, the renormalization must be carried out subtractively instead of 
multiplicatively. The main reason for that is the mishandling of overlapping divergences 
due to the ambiguity inherent in the gauge-technique construction of the vertex, related with the 
unspecified transverse part~\cite{Curtis:1990zs}. 
The (subtractive) renormalization must be carried out at the level of (\ref{Drnfsc}).
Specifically,
\be
\Delta^{-1}_{\s{Q},\s{R}}(q^2)=\frac{Z_A q^2+ i
\left[\widehat\Pi_{\s{Q}}(q^2) + \quark(q^2)\right]}{\left[1+G_{\s{Q}}(q^2)\right]^2},
\label{rgSDE}
\ee 
where
the renormalization constant $Z_A$ is fixed in the MOM scheme through the
condition $\Delta^{-1}_{\s{Q},\s{R}}(\mu^2) = \mu^2$. 
This condition, when applied at the level of Eq.~(\ref{rgSDE}), allows one to express $Z_A$ as
\be
Z_A=[1+G_{\s{Q}}(\mu^2)]^2 - \frac{i}{\mu^2} \left[ \widehat\Pi_{\s{Q}}(\mu^2) + \quark(\mu^2) \right].
\label{z1}
\ee
Now, as is well-known~\cite{Aguilar:2009nf,Aguilar:2009pp},  
the validity of the BRST-driven relation (\ref{funrel}) before and after renormalization prevents  
$G(\mu^2)$ from vanishing when,  according to the MOM prescription, $F(\mu^2) =1$; 
instead, we must impose that $G(\mu^2) =- L(\mu^2)$. 
However, given that  $L(x)$ is considerably smaller 
than $G(x)$ in the entire range of momenta, we can use the approximation $1+G(\mu^2) \approx F^{-1}(\mu^2)=1$,   
without introducing an appreciable numerical error~\cite{Aguilar:2009nf,Aguilar:2009pp}.  Thus,  
we obtain the following approximate equation for $Z_A$ 
\be
Z_A= 1 - \frac{i}{\mu^2} \left[ \widehat\Pi_{\s{Q}}(\mu^2) + \quark(\mu^2) \right]\,,
\label{z12}
\ee
Substituting Eq.~(\ref{z12}) into Eq.~(\ref{rgSDE}), we obtain  
\be
\Delta^{-1}_{\s{Q},\s{R}}(q^2)=\frac{q^2+ i
\left[\widehat\Pi_{\s{Q},\s{R}}(q^2) + \quarkren(q^2)\right]}{\left[1+G_{\s{Q},\s{R}}(q^2)\right]^2},
\label{rgSDE2}
\ee
where, as in (\ref{Xren}), 
 $\widehat\Pi_{\s{Q},\s{R}}(q^2) = \widehat\Pi_{\s{Q}}(q^2)- \frac{q^2}{\mu^2}\widehat\Pi_{\s{Q}}(\mu^2)$, 
while $G_{\s{Q},\s{R}}(q^2) = G_{\s{Q}}(q^2) - G_{\s{Q}}(\mu^2)$.

On the other hand, the exact same procedure yields 
for the renormalized quenched propagator (setting $\quark =0$ and dropping the subscript ``Q'') 
\be
\Delta^{-1}_{\s{R}}(q^2)=\frac{q^2+ i \widehat\Pi_{\s{R}}(q^2)}{\left[1+G_{\s{R}}(q^2)\right]^2}\,.
\label{rgqSDE}
\ee
Then, according to the key operating assumption  explained in the previous section, 
the unquenched quantities $\widehat\Pi_{\s{Q}}(q^2)$ and $G_{\s{Q}}(q^2)$ are to be approximated simply by  
their quenched counterparts, $\widehat\Pi(q^2)$ and $G(q^2)$, respectively.
Consequently, it is easy to verify that 
the renormalized version of (\ref{mastform}) is given by
\be
\Delta_{\s{Q},\s{R}}(q^2) = \frac{\Delta_{\s{R}}(q^2)}
{1 + \left\{ i \,\quarkren(q^2) \left[1+G_{\s{R}}(q^2)\right]^{-2}- \lambda^2 \right\}\Delta_{\s{R}}(q^2)}.
\label{mastformren}
\ee

In this context, the gluon mass related term $\lambda^2$ 
merits some additional comments. As has been emphasized amply in recent works, the seagull identity of (\ref{seagull}), 
when applied to the gluon mass equation, enforces the annihilation of all quadratic 
divergences~\cite{Aguilar:2011ux,Aguilar:2009ke}. 
This is a point of central importance, because 
the disposal of such divergences (had they survived) would require the introduction in the original 
Yang-Mills Lagrangian of a 
counter-term of the form $m^2_0 A^2_{\mu}$, which is, however, forbidden by the local gauge invariance, which  
must remain intact.
Therefore, at least in principle, the renormalization of the gluon mass equation 
proceeds 
as in the case of the homogeneous quark mass equation 
(obtained from the corresponding gap equation without a current mass term), simply 
by renormalizing (multiplicatively) the various quantities appearing on its rhs~\cite{Aguilar:2011ux}.
Note, however, that these considerations, theoretically  important as they may be, 
are of limited practical relevance for the present work, 
because, 
as already mentioned, the  quantity $\lambda^2$
will be not determined dynamically, but rather fitted from the (extrapolated) solutions obtained.

%%%%%%%%%%%%%%%%%%%%%%%%%%%%%%%%%%%%%%%%%%%%%%%%%%%%%%%%%%%%%%%%%%%%%%%%%%%%%%%%%%

\subsection{The transition to Euclidean space}

The actual calculations will be carried out in the Euclidean space, and the 
various relevant formulas, most notably (\ref{mastform}) and (\ref{mastformren}), must be modified accordingly.
In particular, the integral measure is given by
\be
\kint=i\int_{k_\s{\mathrm{E}}}=\frac{i}{(2\pi)^d}
\frac{\pi^{\frac{d-1}2}}{\Gamma\left(\frac{d-1}2\right)}\int_0^\pi\!\diff{}\theta\,\sin^{d-2}\theta\int_0^\infty\!\diff{}y\, y^{\frac d2-1},
\ee 
where $y=k^2$. When $d=4$ this reduces to
\be
\kint=\frac{i}{(2\pi)^3}\int_0^\pi\!\diff{}\theta\sin^{2}\theta\int_0^\infty\!\diff{}y\, y=\frac{i}{(2\pi)^3}\eint \,,
\ee
which is the measure employed in our final results.
In addition, we will use the standard formulas that allow the transition of the various 
Green's functions from the physical Minkowski momentum $q^2$ 
to the Euclidean $q^2_{\s{\mathrm{E}}} = -q^2>0$; specifically
\be
\Delta_\mathrm{\s E}(q^2_\mathrm{\s E})=-\Delta(-q^2_\mathrm{\s E}); \qquad F_\mathrm{\s E}(q^2_\mathrm{\s E})=  F(-q^2_\mathrm{\s E});
\qquad G_\mathrm{\s E}(q^2_\mathrm{\s E})= G(-q^2_\mathrm{\s E}). 
\label{euc1}
\ee
and
\be
A_\mathrm{\s E}(q^2_\mathrm{\s E})=  A(-q^2_\mathrm{\s E}) \qquad B_\mathrm{\s E}(q^2_\mathrm{\s E})=  B(-q^2_\mathrm{\s E})\,.
\label{euc2}
\ee
The Euclidean version of $\quark(q^2)$  is defined as the 
result of the aforementioned operations 
at the level
of (\ref{quark-loop-full}), but with the imaginary factor $i$ that comes 
from the measure absorbed by the factor of $i$ multiplying $\quark(q^2)$  in Eqs.~(\ref{mastform}) or (\ref{mastformren}).
Effectively, this amounts to the substitution
$i\quark(q^2)\to - \widehat{X}_\mathrm{\s E}(q^2_\mathrm{\s E})$
where the $\widehat{X}_\mathrm{\s E}(q^2_\mathrm{\s E})$ is obtained from (\ref{quark-loop-full}) by replacing $\kint \to \int_{k_\s{\mathrm{E}}}$, (no more $i$)
euclidianizing the momenta ($q^2 \to -q^2_\mathrm{\s E}$, $k^2 \to -k^2_\mathrm{\s E}$), and using (\ref{euc2}).
Then, the euclidian version of (\ref{mastform}) becomes (we suppress the suffix ``E'' throughout)
\be
\Delta_{\s{Q}}(q^2) = \frac{\Delta(q^2)}
{1 + \left\{ \quark(q^2) \left[1+G(q^2)\right]^{-2}+ \lambda^2 \right\}\Delta(q^2)}.
\label{mastformeuc}
\ee
The conversion of (\ref{mastformren}) to Euclidean space proceeds following exactly analogous steps. 

One may carry out two elementary checks of the expression given in (\ref{mastformeuc}).
First, in the IR limit, $q^2=0$, after using (\ref{quark-loop-full-0}), $\Delta^{-1}(0) =m^2(0)$, and the definition of 
$\lambda^2$ in (\ref{lambda}), we obtain $\Delta^{-1}_{\s{Q}}(0)= m^2_{\s{Q}}(0)$, as we should.

In the opposite limit, where $q^2$ acquires large values compared to all mass scales involved, 
we substitute into \1eq{mastformeuc} the perturbative one-loop results, keeping terms 
up to order $\alpha_s$. 
The  Euclidean version of (\ref{qr1}) is determined following the steps described above; specifically, since  
\be
i\quark^{[1]}(q^2) = - \frac{\alpha_s}{6\pi} q^2 \ln (-q^2/\mu^2) \,,
\ee 
then (restoring the ``E'' for this step only)
\bea 
\quark^{[1]}_{\mathrm{\s E}}(q^2_{\mathrm{\s E}}) &=&  \left\{\frac{\alpha_s}{6\pi} q^2 \ln (-q^2/\mu^2) \right\}_{q^2\to -q_{\mathrm{\s E}}^2}
\\\nonumber
&=& - \frac{\alpha_s}{6\pi} q^2_{\mathrm{\s E}} \ln (q^2_{\mathrm{\s E}}/\mu^2) \,.
\eea
Combining this with the standard result  
\be
[\Delta^{-1}(q^2)]^{[1]}=  q^2 \left[1+  \frac{13 C_A \alpha_s}{24 \pi}\ln (q^2 /\mu^2)\right]\,,
\label{1loop_g}
\ee 
we obtain from (\ref{mastformeuc}) [with $n_f$ quark flavors, and $C_A=3$] 
\be
[\Delta^{-1}_{\s{Q}}(q^2)]^{[1]} = 
q^2 \left[1 +  \frac{\alpha_s}{8\pi} \left\{ 13 - \frac{4}{3} n_f\right\} \ln (q^2 /\mu^2)\right]\,,
\label{pert_unq}
\ee
which is the correct one-loop result (in the Landau gauge). 

In the derivation given above, the perturbative expression for $G$, namely ($C_A=3$)
\be
1+G^{[1]}(q^2) = 1 + \frac{9\alpha_s}{16\pi}  \ln (q^2 /\mu^2)\,,
\label{pertG}
\ee
was not necessary, since, its inclusion in \1eq{mastformeuc} gives contributions of ${\cal O}(g^4)$;
however, (\ref{pertG}) is needed for a final check. Specifically, as 
is well known, 
due to the QED-like WIs characteristic of PT-BFM scheme,
the PT-BFM propagator, usually denoted by $\widehat\Delta$, 
captures the running  of the gauge coupling ($\beta$ function),
for any value of the gauge-fixing parameter.
$\widehat\Delta$ and $\Delta$ are related by the all-order relation~\cite{Binosi:2002ft,Grassi:1999tp} 
\be
\widehat{\Delta}^{-1}(q^2)=
\left[1+G(q^2)\right]^2 \Delta^{-1}(q^2)\,,
\label{bqi}
\ee
whose perturbative expansion yields 
\be
[\widehat{\Delta}^{-1}_{\s{Q}}(q^2)]^{[1]} = 
q^2 \left[1 +  \frac{\alpha_s}{48\pi} \left\{ 33 - 2 n_f\right\} \ln (q^2 /\mu^2)\right]\,,
\ee
namely the correct one-loop result.

%%%%%%%%%%%%%%%%%%%%%%%%%%%%%%%%%%%%%%%%%%%%%%%%%%%%%%%%%%

\section{\label{numres}Numerical results}

In this section we will first review the lattice 
data for the quenched gluon propagator $\Delta(q^2)$ and ghost dressing function $F(q^2)$,
and the nonperturbative expressions of the quark functions $A(p^2)$ and $B(p^2)$, 
obtained in~\cite{Aguilar:2010cn} from the solution of the quark gap equation. 
With all necessary ingredients  available, \ie \mbox{$\Delta(q^2)$, $F(q^2)$, $A(q^2)$ and $B(q^2)$},
we then evaluate numerically the integrals that determine the contribution
of the quark loop, $\quark(q^2)$ (BC vertex), and  $\quark^{\s{\rm CP}}(q^2)$ (CP vertex), given 
by Eqs.~(\ref{quark-loop-full}) and ~(\ref{delta_cp}), respectively. 
Finally, with the quark loop contribution at our disposal, we proceed to estimate
through Eq.~(\ref{mastformeuc}) the effect of ``unquenching'', namely     
how the overall shape of the quenched propagator $\Delta(q^2)$  
is affected by the presence of the quark loops. 
Finally, we compare the resulting dressing function with that obtained from unquenched lattice simulations.
Given the amount of information presented in this section, 
we have organized the material in four subsections, and
have enumerated the main points of each subsection, to facilitate the perusal.

\subsection{Ingredients}

\n{i} The starting point of our  numerical analysis are the quenched $SU(3)$
lattice  results  for the  gluon  propagator  $\Delta(q^2)$ and  ghost
dressing  function $F(q^2)$~\cite{Bogolubsky:2007ud}. These  are shown,
respectively, on the left  and right panels of Fig.~\ref{fig:prop}, for
three  different renormalization  points \mbox{({$\mu=4.3$,  $3.0$ and
$2.3$ GeV)}}. On  the same figure we also  plot the corresponding fits
for  the  three  different  renormalization  points;  the explicit 
functional form used for these fits can be
found in various recent articles~\cite{Aguilar:2010cn,Aguilar:2011ux,Aguilar:2010gm}.

%%%%%%%%%%%%%%%%%%%%%%%%%%%%%%%%%%%%%%%%%%%%%%%%%%%%%%%%%%%%%%%%%%%%%%%%%%
%             Fig.5  gluon propagator and ghost dressing SU(3)
%%%%%%%%%%%%%%%%%%%%%%%%%%%%%%%%%%%%%%%%%%%%%%%%%%%%%%%%%%%%%%%%%%%%%%%%%%%%
\begin{figure}[!t]
%\hspace{.1cm}
\begin{minipage}[b]{0.45\linewidth}
%\centering
\includegraphics[scale=0.5]{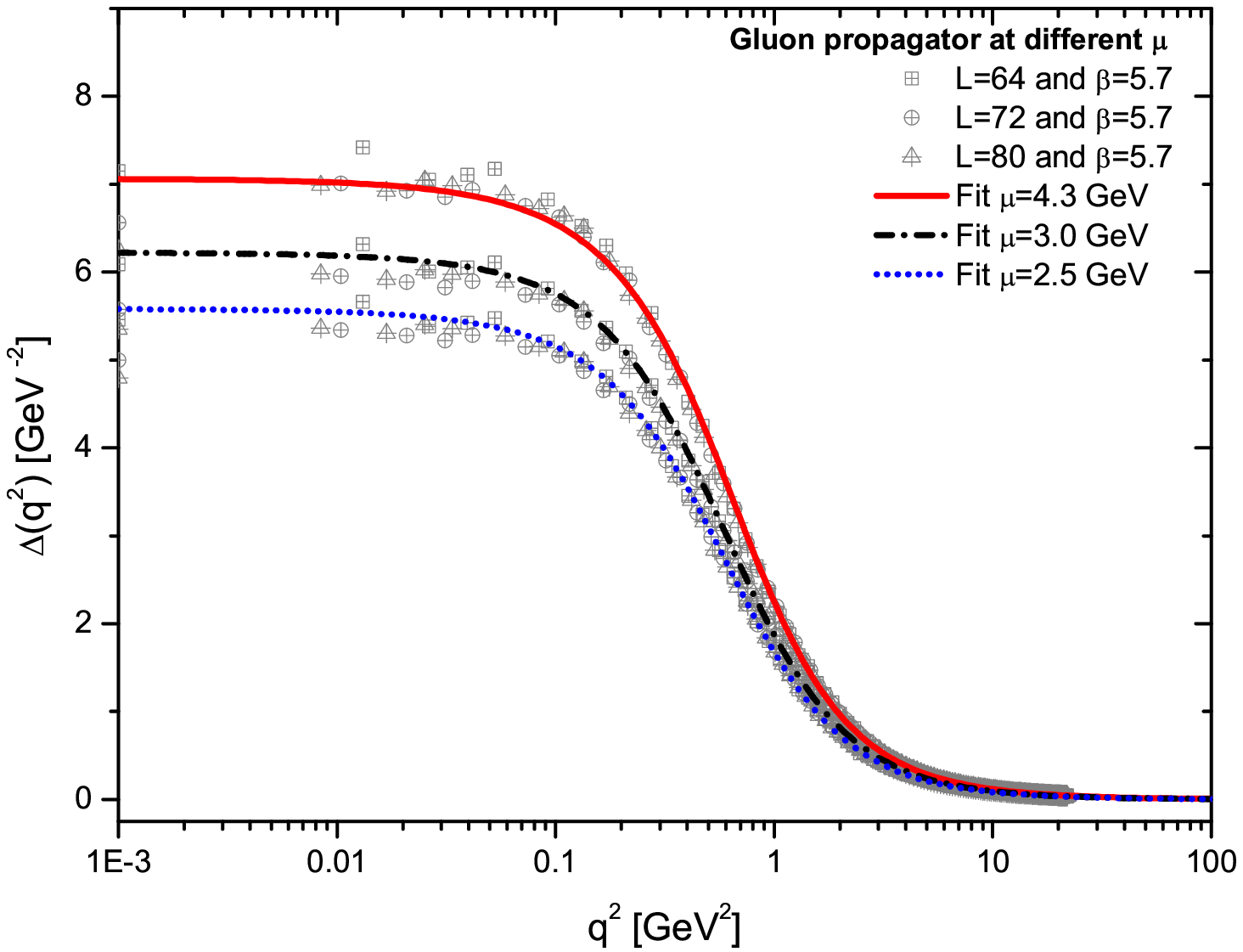}
\end{minipage}
\hspace{0.5cm}
\begin{minipage}[b]{0.50\linewidth}
\hspace{-1cm}
\includegraphics[scale=0.51]{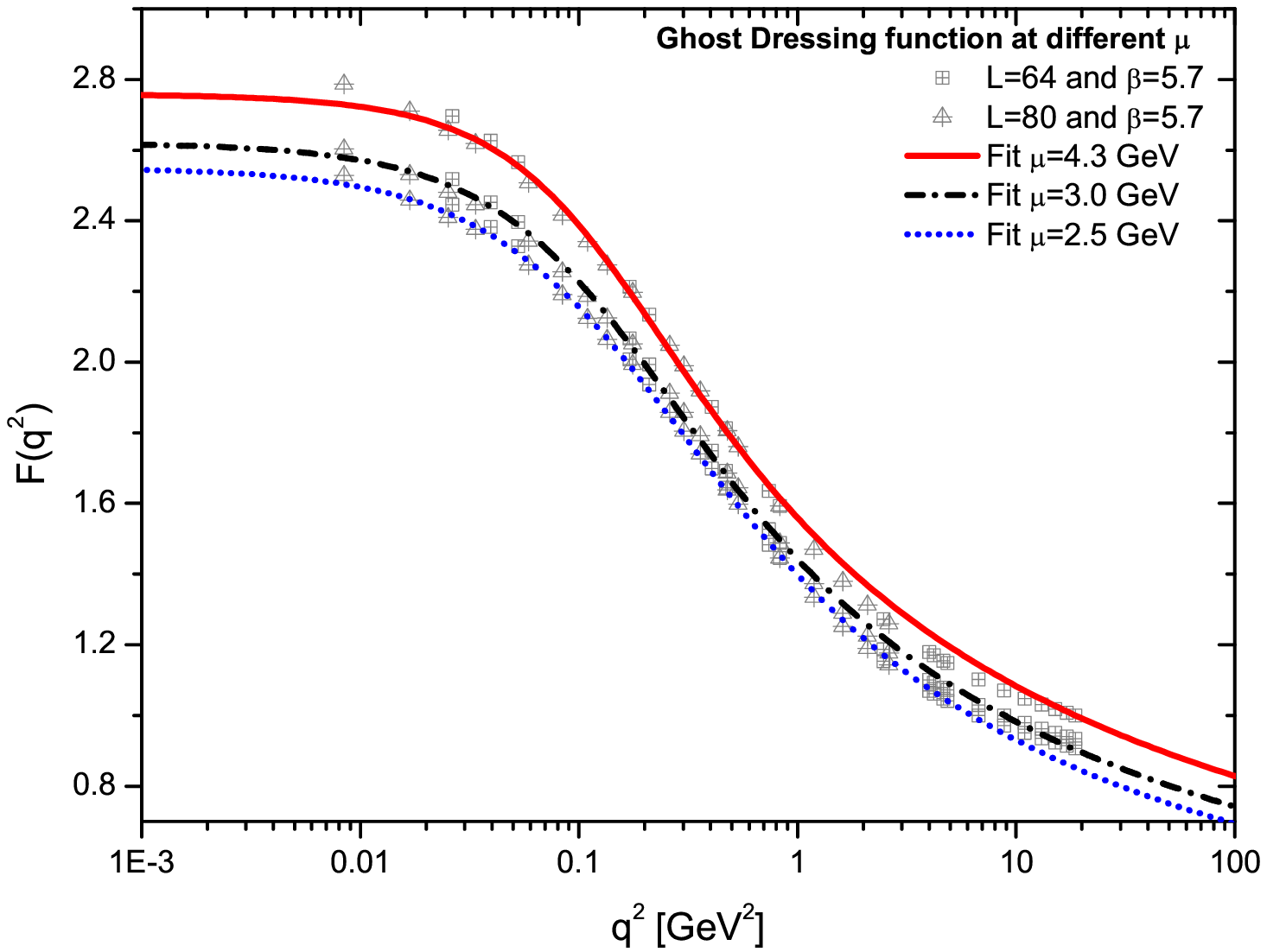}
\end{minipage}
\vspace{-0.75cm}
\caption{\label{fig:prop}Lattice result for the $SU(3)$ gluon 
propagator (left panel) and ghost dressing function (right panel) renormalized  
at  three different points: \mbox{\mbox{$\mu=4.3$ GeV}} 
(solid red curve), \mbox{\mbox{$\mu=3.0$ GeV} (dash-dotted black curve)}, and 
\mbox{\mbox{$\mu=2.5$ GeV} (dotted blue curve)}.}
\end{figure}
%%%%%%%%%%%%%%%%%%%%%%%%%%%%%%%%%%%%%%%%%%%%%%%%%%%%%%%%%%%%%%%%%%

\n{ii} Next,  the computation  of  the quark  contribution $\quark(q^2)$  and
$\quarkCP(q^2)$  [\2eqs{quark-loop-full}{delta_cp}] requires  the
knowledge of  the nonperturbative behavior of  the functions $A(k^2)$
and  $B(k^2)$   appearing  in  the   definition  of  the   full  quark
propagator~\noeq{qprop}. Both  functions can be  determined by solving
numerically the quark  gap equation; however, one has  to be particular
careful on how the non-Abelian  quark-gluon vertex, which enters in the
latter  equation, is  approximated. Note in particular that, 
as  discussed in  detail  in~\cite{Aguilar:2010cn},
the quark gap equation is identical within both the 
conventional and the PT-BFM frameworks. 
As a result, the quark-gluon vertex entering in it is $\G_\mu$  
(and not $\widehat\G_{\mu}$), satisfying the STI given in \1eq{STI}. 
This fact, in turn, 
introduces a numerically crucial  dependence on the
ghost dressing function and the quark-ghost scattering amplitude. Once
these  effects are  duly taken  into account,  and the  BC or  CP vertices
improved   accordingly~\cite{Aguilar:2010cn},   one   can  solve   the
resulting  nonlinear system  of  integral equations  for $A(k^2)$  and
$B(k^2)$,  supplemented by the  lattice gluon  propagator and  ghost
dressing function mentioned above.

\n{iii} The results obtained following the outlined procedure are shown in Fig.~\ref{fig:AB} (for the specific value $\mu = 4.3$ GeV in this case).  In particular, on the left panel we plot the inverse of the quark wave function $A^{-1}(k^2)$ for the improved BC vertex (dotted black curve), and the  ``improved'' CP vertex (dashed blue curve);
on the right panel we show the corresponding solutions for the $B(k^2)$ function. 
At this point the momentum dependence 
of the dynamical quark mass ${\mathcal M}(k^2)$ can be straightforwardly obtained, since  \mbox{$\qm(k^2)=B(k^2)/A(k^2)$}, 
and is plotted in Fig.~\ref{fig:dymass}, for the two forms of the quark gluon vertex considered. 
Clearly the two results coincide in the UV, 
whereas in the IR we notice that  the CP vertex produces the slightly higher value 
\mbox{${\mathcal M}(0)=M=307$} MeV when compared with the BC vertex result \mbox{$M=292$ MeV}.
Note, finally, that the results presented have been obtained  
in the chiral limit, where no ``current'' mass has been used when solving the gap equation.  
 
%
%%%%%%%%%%%%%%%%%%%%%%%%%%%%%%%%%%%%%%%%%%%%%%%%%%%%%%%%%%%%%%%%%%%%%%%%%%
%             Fig.6  solutions for A and B
%%%%%%%%%%%%%%%%%%%%%%%%%%%%%%%%%%%%%%%%%%%%%%%%%%%%%%%%%%%%%%%%%%%%%%%%%%%%
\begin{figure}[!t]
%\hspace{-1.5cm}
\begin{minipage}[b]{0.45\linewidth}
\centering
\includegraphics[scale=0.5]{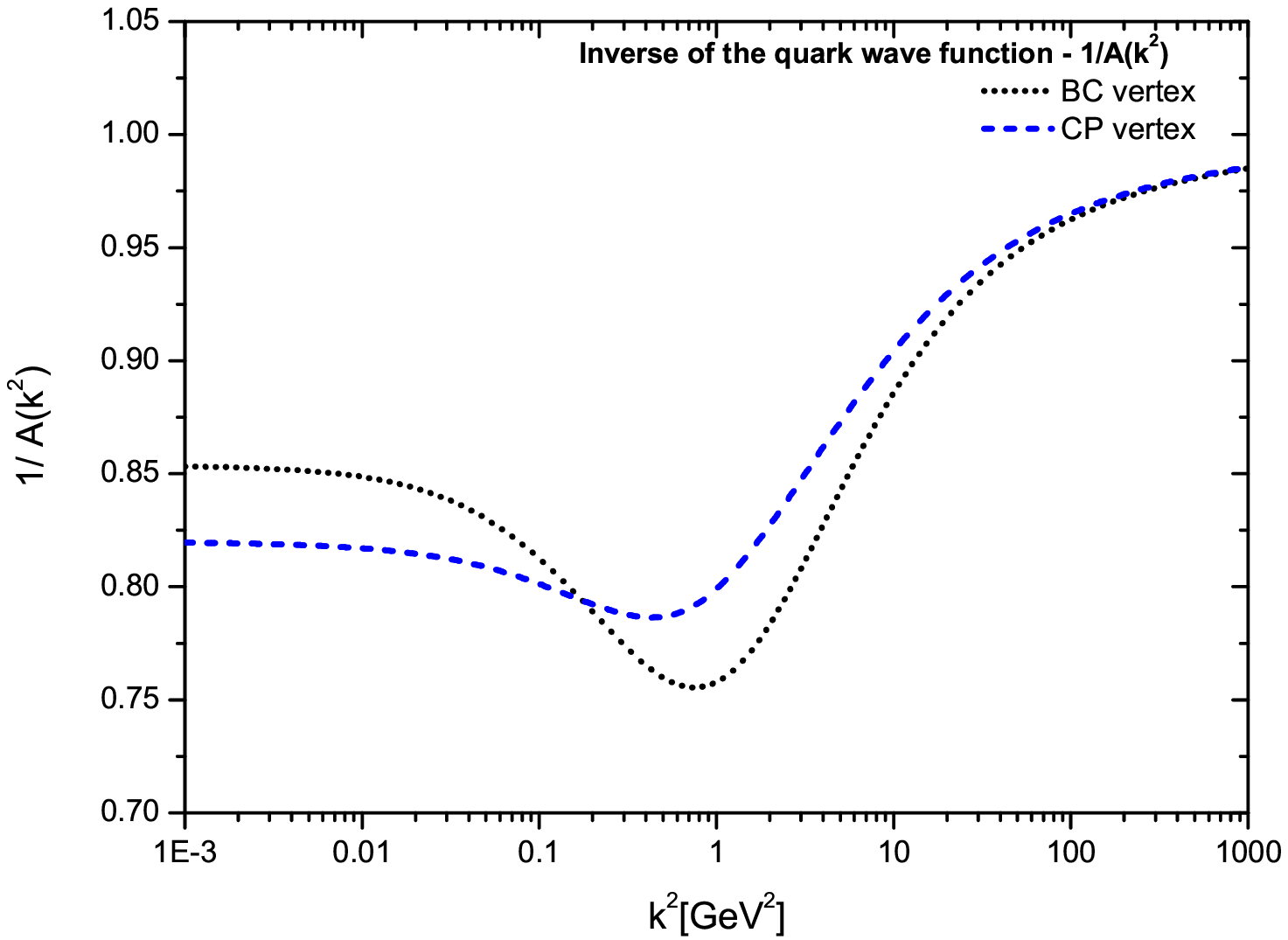}
\end{minipage}
\hspace{0.5cm}
\begin{minipage}[b]{0.50\linewidth}
\hspace{-1.5cm}
\includegraphics[scale=0.5]{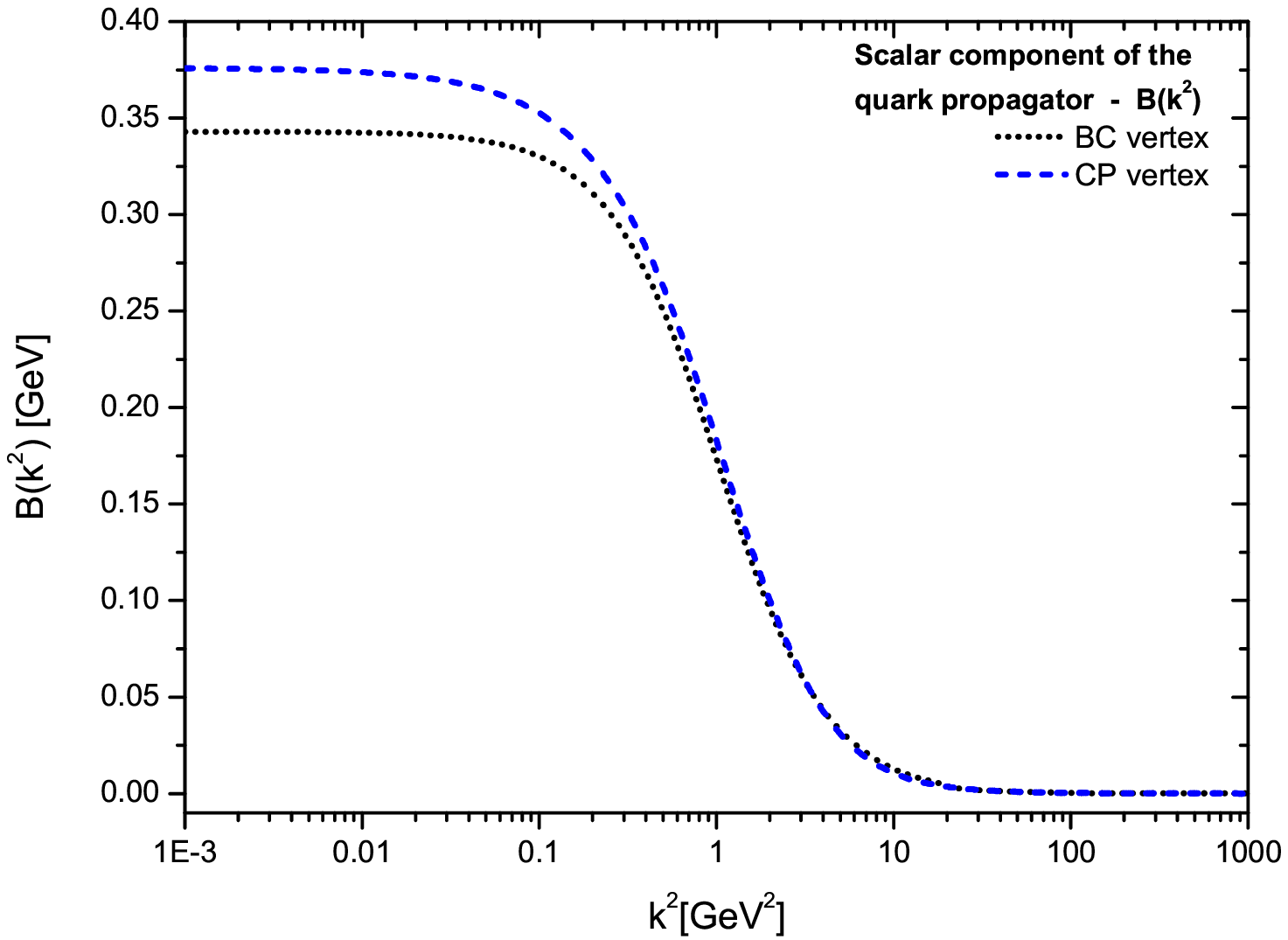}
\end{minipage}
\vspace{-0.75cm}
\caption{\label{fig:AB}Solution of the quark gap equation:  $A^{-1}(k^2)$ (left panel) and $B(k^2)$ (right panel) renormalized at $\mu= 4.3$ GeV. Dotted black curves correspond to the improved BC vertex, while dashed blue curves to the improved CP vertex.}
\end{figure}
%%%%%%%%%%%%%%%%%%%%%%%%%%%%%%%%%%%%%%%%%%%%%%%%%%%%%%%%%%%%%%%%%%%% 

%%%%%%%%%%%%%%%%%%%%%%%%Fig.7 %%%%%%%%%%%%%%%%%%%%%%%%%%%%%%%%%%%%%
%            Fig. 7  - dynamical mass
%%%%%%%%%%%%%%%%%%%%%%%%%%%%%%%%%%%%%%%%%%%%%%%%%%%%%%%%%%%%%%%%%%%
\begin{center}
\begin{figure}[!t]
\includegraphics[scale=0.5]{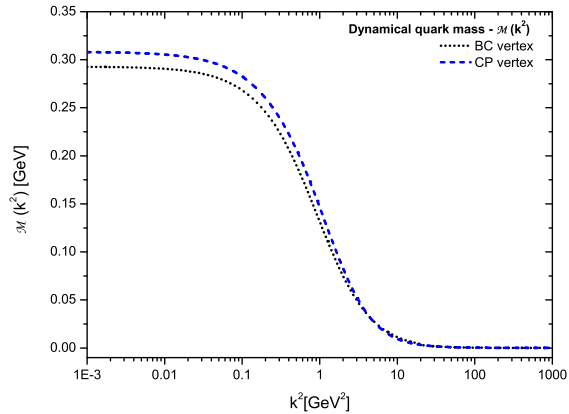}
\caption{The momentum dependence of the dynamical quark mass \mbox{${\mathcal M}(k^2)=B(k^2)/A(k^2)$},
using the same conventions as in the previous plot.} 
\label{fig:dymass}
\end{figure}
\end{center}
%%%%%%%%%%%%%%%%%%%%%%%%%%%%%%%%%%%%%%%%%%%%%%%%%%%%%%%%%%%%%%%%%%% 

\subsection{The quark loop}

%%%%%%%%%%%%%%%%%%%%%%%%%%%%%%%%%%%%%%%%%%%%%%%%%%%%%%%%%%%%%%%%%%%%%%%%%
%             Fig.8  T contributions
%%%%%%%%%%%%%%%%%%%%%%%%%%%%%%%%%%%%%%%%%%%%%%%%%%%%%%%%%%%%%%%%%%%%%%%%%%%%
\begin{figure}[!t]
%\hspace{-1.5cm}
\begin{minipage}[b]{0.45\linewidth}
\includegraphics[scale=0.5]{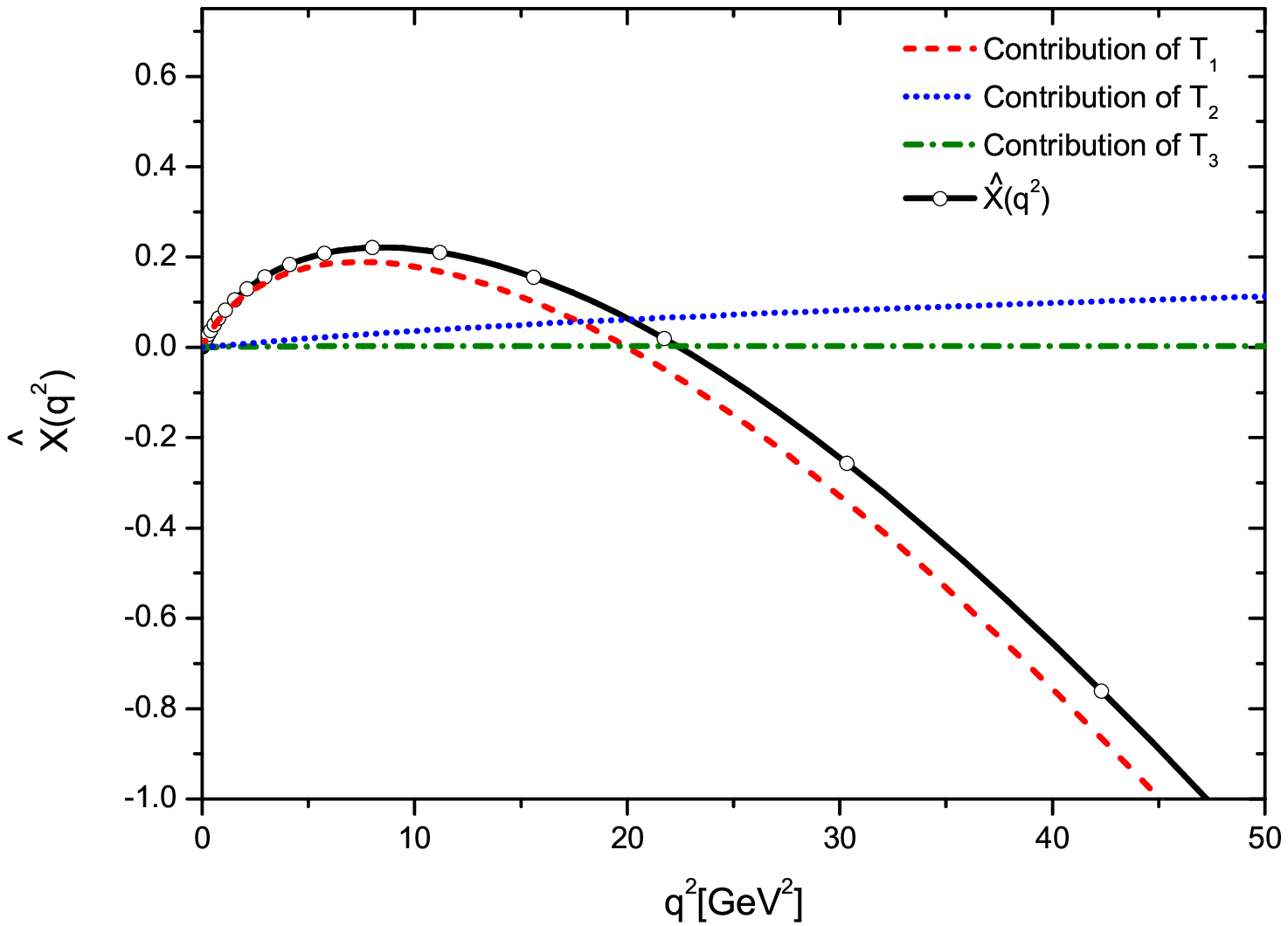}
\end{minipage}
\hspace{0.5cm}
\begin{minipage}[b]{0.50\linewidth}
\hspace{-1.5cm}
\includegraphics[scale=0.5]{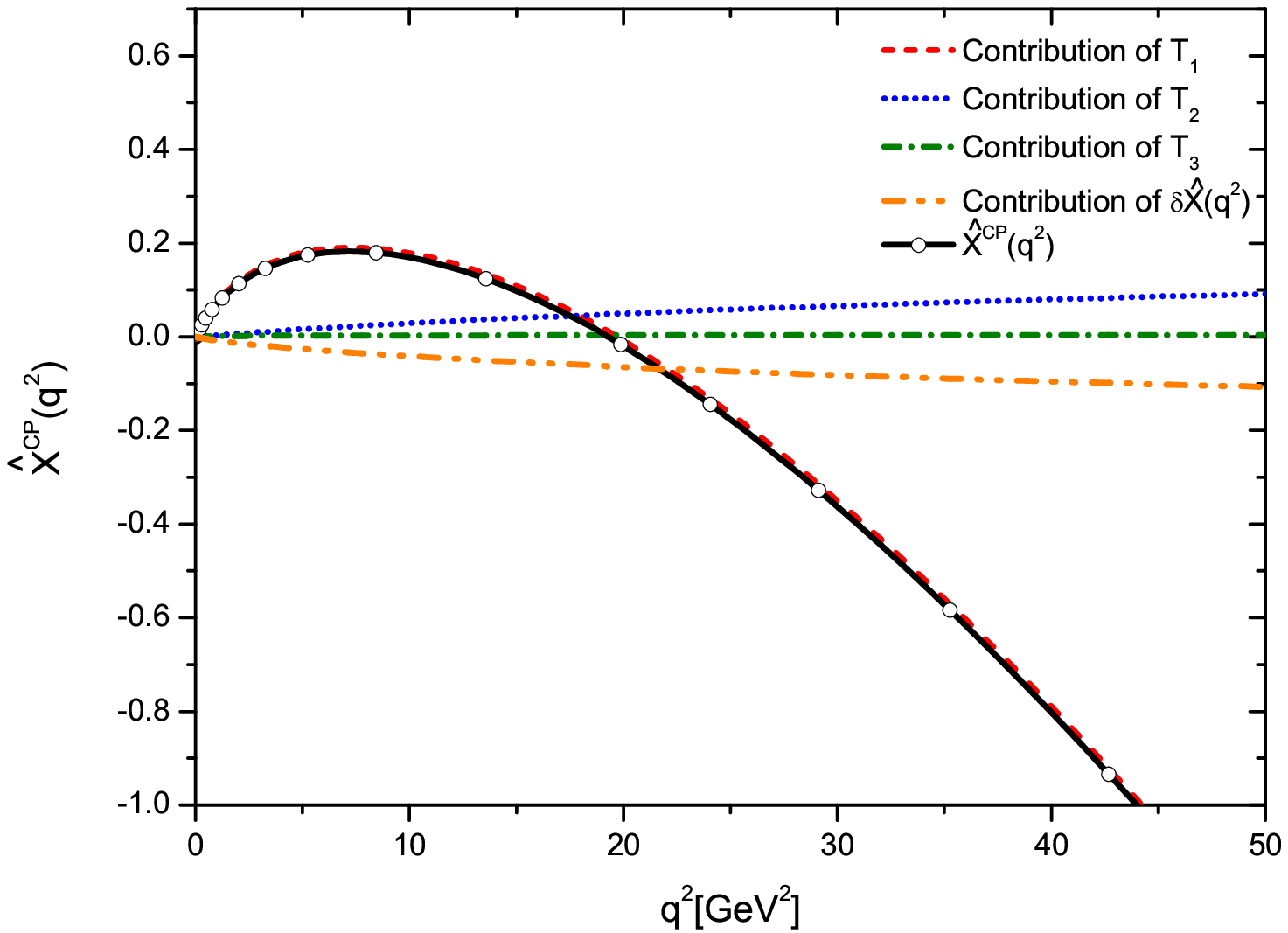}
\end{minipage}
\vspace{-0.75cm}
\caption{\label{fig:resx}Individual contributions of the terms  proportional to $T_1$ (dashed red curve), $T_2$ (dotted blue line),  $T_3$ (dash-dotted green) and $\delta\quark(q^2)$ (dashed with two dots orange curve), to  $\quark(q^2)$ (left panel) and $\quarkCP(q^2)$ (right panel) respectively. The sum of all contributions produce in both cases the continuous (black with white circles) curve, which represent the full quark loop contribution to the gluon propagator.}
\end{figure}

We can now proceed to the numerical evaluation 
of the full quark loop, namely $\quark(q^2)$ (BC vertex) and  $\quarkCP(q^2)$ (CP vertex), as given by  \2eqs{quark-loop-full}{delta_cp} respectively. 

\n{i} On the left panel of  Fig.~\ref{fig:resx} we show the results obtained
for each individual contribution of $\quark(q^2)$, as expressed in \2eqs{quark-loop-full}{T123}. 
As can be easily seen, the leading contribution comes from the $T_1$ term, which, as shown in the Appendix, is also    
the term responsible for the appearance of the perturbative logarithm; the $T_2$ and $T_3$ contributions are instead subdominant.

\n{ii} On the right panel of Fig.~\ref{fig:resx}, we show the same quantities 
for the $\quarkCP(q^2)$ term. In this case, one has the additional contribution $\delta \quark(q^2)$, given in~\1eq{quarkCP}, 
coming from the inclusion of the transverse part of the quark-gluon vertex. 
The net numerical effect is that the latter term will almost completely cancel the subdominant 
terms $T_2$ and $T_3$, so that the $T_1$ term practically coincides with the full answer.

\n{iii} The results for the quark loop $\quark(q^2)$ and $\quarkCP(q^2)$ are finally compared in \fig{fig:pi_bc_cp} for the $n_f=2$ case.  It is important to notice that, indeed, $\quark(0)=\quarkCP(0)= 0$, as we had 
previously announced.

%%%%%%%%%%%%%%%%%%%%%%%%%%%%%%%%%%%%%%%%%%%%%%%%%%%%%%%%%%%%
%         Fig 9 X comparison
%%%%%%%%%%%%%%%%%%%%%%%%%%%%%%%%%%%%%%%%%%%%%%%%%%%%%%%%%%%%%%%%%%%
\begin{center}
\begin{figure}[!t]
\includegraphics[scale=0.5]{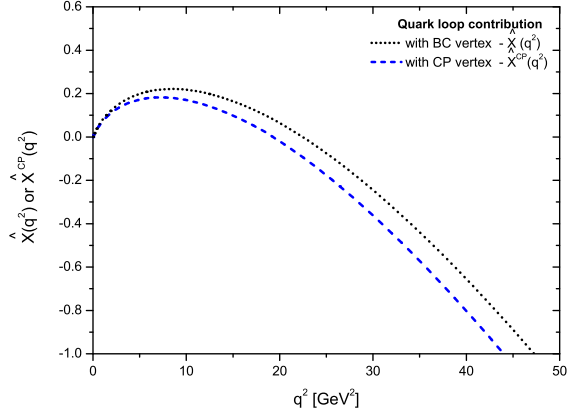}
\caption{Comparison between the contributions of the quark loop $\quark(q^2)$ (dotted black curve) and $\quarkCP(q^2)$ (dashed blue curve) to the gluon self-energy with $n_f=2$.} 
\label{fig:pi_bc_cp}
\end{figure}
\end{center}
%%%%%%%%%%%%%%%%%%%%%%%%%%%%%%%%%%%%%%%%%%%%%%%%%%%%%%%%%%%%%%%%%%% 
%%%%%%%%%%%%%%%%%%%%%%%%%%%

\subsection{Effect on the gluon propagator}

\n{i} The next step is to compute the unquenched gluon propagator given in \1eq{mastformeuc}. 
The first thing we should notice is the presence of the auxiliary function $1+G(q^2)$ in
the denominator of \1eq{mastformeuc}. Using the  fact that, in the Landau gauge, $L(q^2)$
is numerically suppressed~\cite{Aguilar:2009nf,Aguilar:2009pp},  it follows immediately  
from \1eq{funrel} that \mbox{$1+G(q^2) \approx  F^{-1}(q^2)$}.  

\n{ii} Substituting  into \1eq{mastformeuc} the results for  $\Delta(q^2)$ and $F(q^2)$, 
renormalized at \mbox{$\mu = 4.3$ GeV} and  presented in Fig.~\ref{fig:prop}, together
with either $\quark(q^2)$ (BC vertex) or  $\quarkCP(q^2)$ (CP vertex)
of \fig{fig:pi_bc_cp}, we obtain the results shown on the left panel of Fig.~\ref{fig:unq}. 
As before, the  dotted black curve represents the result for the case where we employ  the BC vertex,  while the 
dashed blue curve is for the CP vertex. We clearly see that the unquenched gluon propagator suffers  a sizable
suppression in the intermediate momenta region compared to the quenched case (solid red curve). 
Notice that, in this particular case (left panel of Fig.~\ref{fig:unq}) we have set $\lambda^2 = 0$, and therefore,  
the three curves coincides 
at $q^2=0$ since $\quark(0)=\quarkCP(0)= 0$.  

\n{iii} As mentioned before, due to our present limitation in determining the precise value of $\lambda^2$, 
we will restrict ourselves to extracting an approximate range for $\lambda^2$, through the extrapolation
of the curves in the region delimited by the shaded area showed on the left panel of Fig.~\ref{fig:unq}.

%%%%%%%%%%%%%%%%%%%%%%%%%%%%%%%%%%%%%%%%%%%%%%%%%%%%%%%%%%%%%%%%%%%%%%%%%
%             Fig.10  gluon propagator BC and CP -band and extrapolation
%%%%%%%%%%%%%%%%%%%%%%%%%%%%%%%%%%%%%%%%%%%%%%%%%%%%%%%%%%%%%%%%%%%%%%%%%%%%
\begin{figure}[!t]
%\hspace{-1.5cm}
\begin{minipage}[b]{0.45\linewidth}
\includegraphics[scale=0.5]{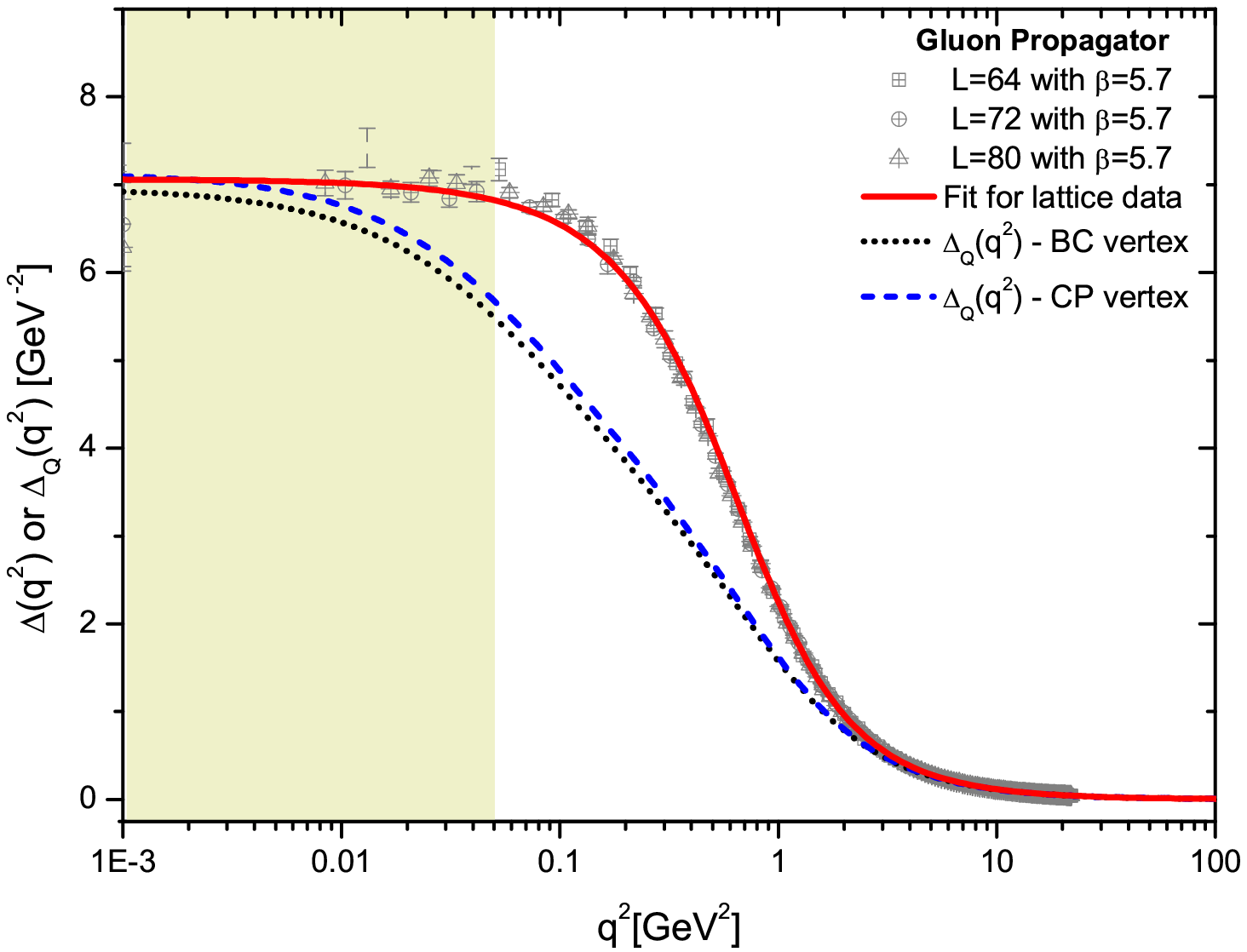}
\end{minipage}
\hspace{0.5cm}
\begin{minipage}[b]{0.50\linewidth}
\hspace{-1.5cm}
\includegraphics[scale=0.5]{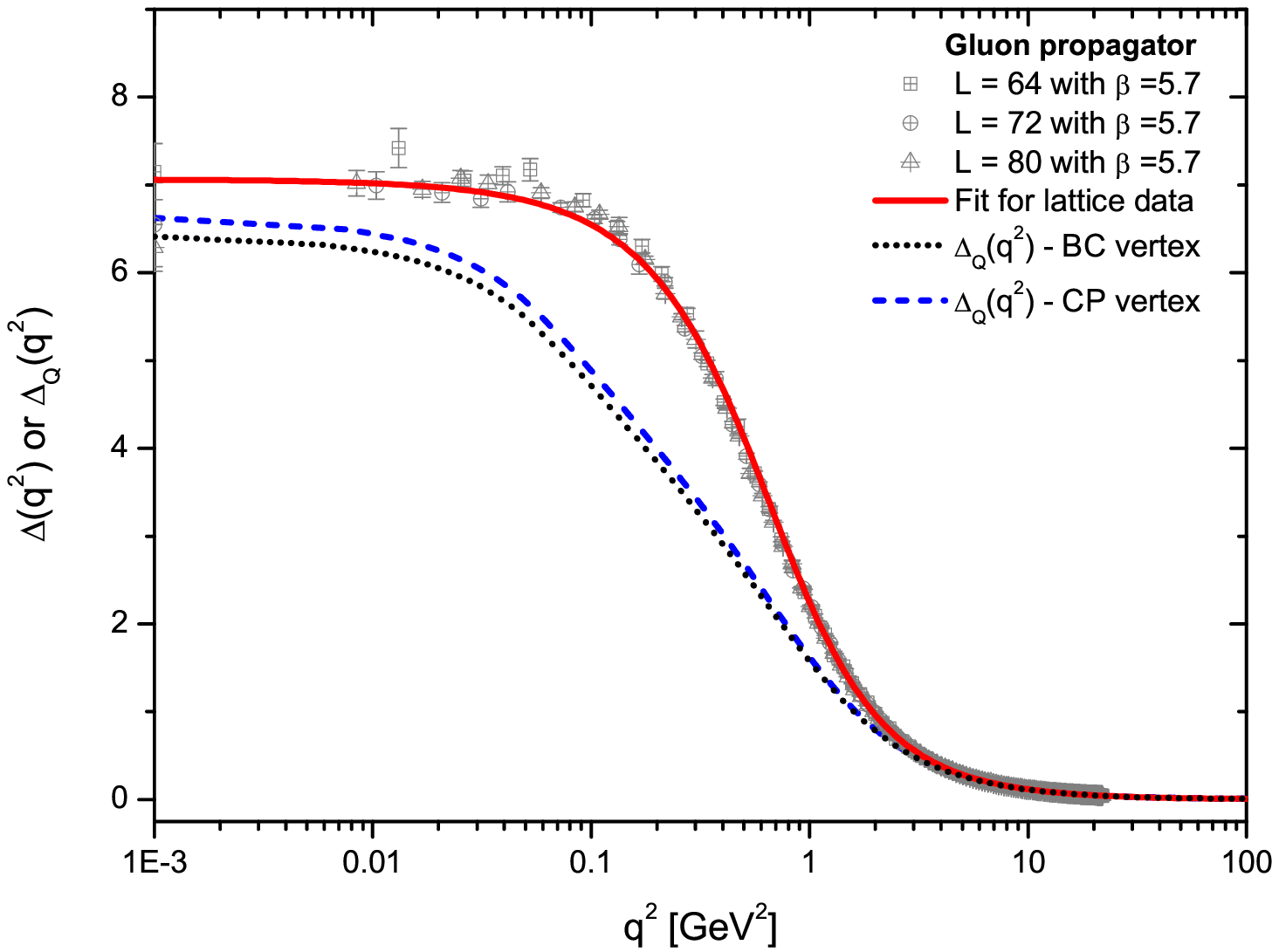}
\end{minipage}
\vspace{-0.75cm}
\caption{\label{fig:unq} The unquenched gluon propagator $\Delta_Q(q^2)$  when no extrapolation is used, i.e. $\lambda^2=0$ in
\1eq{mastformeuc} (left panel). The dotted black curve represents the unquenched propagator obtained with the BC vertex whereas 
the dashed blue curve represents the result for the CP vertex. The result for $\Delta_{\s Q}(q^2)$  when the  extrapolation is performed 
in the shade area i.e. from $q^2 = 0.05\,\mbox{GeV}^2$ towards the deep IR (right panel).}
\end{figure}

%%%%%%%%%%%%%%%%%%%%%%%%%%%%%%%%%%%%%%%%%%%%%%%%%%%%%%%%%%%%%%%%%%%%

Specifically, 
we perform a one-dimensional
extrapolation in the deep IR region using as input
the result obtained for the quenched gluon propagator in the middle IR
and intermediate regions.
The first step is to select the momentum from which
$\Delta_Q(q)$ is extrapolated.
We choose three different points, namely \n{i} \mbox{$0.02\,\mbox{GeV}^2$},
\n{ii} \mbox{$0.05\,\mbox{GeV}^2$} and \n{iii} \mbox{$0.07\,\mbox{GeV}^2$},
and implement the extrapolation starting for each of these points.
In all three cases, we extrapolate the data up to $q^2=10^{-3}
\,\mbox{GeV}^2$  using
the cubic B-spline method. We basically split each of these ranges
into a 150  pieces, and fit each segment with a cubic
Bezier spline. The goal is to get a fit segment that is smooth in the
first derivative, and continuous in
the second derivative, both within an interval and at its boundaries.
When these boundary conditions are met, the entire function is constructed
in a piece-wise manner.

On the right panel of Fig.~\ref{fig:unq}, we show $\Delta_Q(q^2)$  when 
the extrapolation is done for values of momenta smaller than \mbox{$q^2=0.05\,\mbox{GeV}^2$}. 
As we can clearly see, the tendency of the unquenched gluon propagator 
is always to be below the quenched one (solid red curve), no matter
if we use the BC vertex (dotted black curve) or the CP vertex (dashed blue curve). 

Now, we are in position to determine the order of magnitude of~$m^2_{\s Q}(0)$ and $\lambda^2$.
Combining \2eqs{h2}{lambda} and the data presented
on the right panel of  Fig.~\ref{fig:unq},  we found the values of \mbox{$m^2_{\s Q}(0)= 0.156 \,\mbox{GeV}^2$}, 
\mbox{$m^2(0)= 0.142 \,\mbox{GeV}^2$} and $\lambda^2 = 0.014\,\mbox{GeV}^2$ for the BC vertex, whereas for the CP
we have \mbox{$m^2_{\s Q}(0)= 0.151\,\mbox{GeV}^2$} and $\lambda^2 = 0.009\,\mbox{GeV}^2$. 
These results suggest that the effective gluon mass increases when we include the quark loops in the gluon
self-energy. 

%%%%%%%%%%%%%%%%%%%%%%%%%%%%%%%%%%%%%%%%%%%%%%%%%%%%%%%%%%%%%
%         Fig. 11 extrapolation
%%%%%%%%%%%%%%%%%%%%%%%%%%%%%%%%%%%%%%%%%%%%%%%%%%%%%%%%%%%%%%%%%%%
\begin{center}
\begin{figure}[!t]
\includegraphics[scale=0.5]{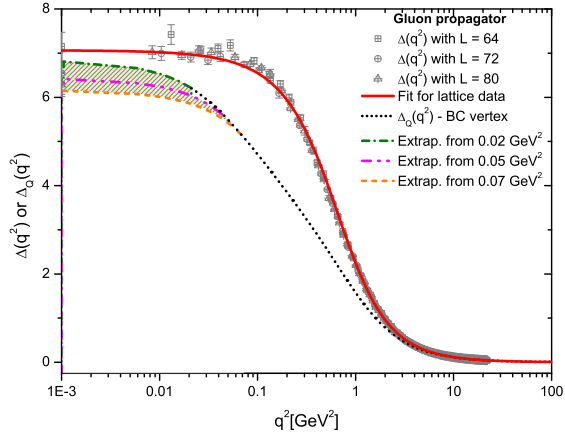}
\caption{\label{fig:band} Comparison between the quenched $\Delta(q^2)$ and the unquenched $\Delta_{\s Q}(q^2)$ gluon propagators. The yellow striped band shows
the possible values that $\Delta_{\s Q}(0)$ can assume at zero momentum. The two
curves delimiting the band represent the extrapolation towards the IR either
starting  from $0.02\, \mbox{GeV}^2$ (dash-dotted green line) or  $0.07\, \mbox{GeV}^2$ (dashed orange line). 
The dashed with two dots magenta curve corresponds to an extrapolation of the numerical result starting from the intermediate value $0.05\,\mbox{GeV}^2$.}
\end{figure}
\end{center}
%%%%%%%%%%%%%%%%%%%%%%%%%%%%%%%%%%%%%%%%%%%%%%%%%%%%%%%%%%%%%%%%%%% 

In addition,  notice that the results obtained
with the BC and CP vertices differ only by approximately $3\%$.  Since
this difference is rather small and does not cause significant changes
in what  follows, for the  rest of  our analysis we will focus on  the BC
vertex only.

\n{iv}
It is important to verify whether the
IR suppression, shown in the  unquenched propagator of the right panel of Fig.~\ref{fig:unq},
persists when we start the curve extrapolation from different values.  This is shown in Fig.~\ref{fig:band}, where we compare the results obtained 
when we  extrapolate $\Delta_{\s Q}(q^2)$ (with the BC vertex) from momenta below $q^2 = 0.02\,\mbox{GeV}^2$ (dash-dotted green line),
$q^2 = 0.05\,\mbox{GeV}^2$ (dashed  with two dots magenta line), and $q^2 = 0.07\,\mbox{GeV}^2$ (dashed orange line).  

Indeed, we can see that the general trend, for all cases, is that the unquenched propagator $\Delta_{\s Q}(q^2)$  displays 
suppressed intermediate and IR regions, when compared to the quenched case. 
In particular, for the extrapolation starting at $q^2 = 0.02\,\mbox{GeV}^2$
we can see that \mbox{$m^2_{\s Q}(0)= 0.147 \,\mbox{GeV}^2$} and \mbox{$\lambda^2 = 0.005\,\mbox{GeV}^2$}; whereas
when we extrapolate from  $q^2 = 0.07\,\mbox{GeV}^2$ we obtain 
\mbox{$m^2_{\s Q}(0)= 0.163 \,\mbox{GeV}^2$} and \mbox{$\lambda^2 = 0.021\,\mbox{GeV}^2$}.
Therefore, the extrapolations mentioned above produce  
a range of  possible values for $\Delta_{\s Q}(0)$ or, equivalently, $m^2_{\s Q}(0)$ indicated by the yellow 
striped band on Fig.~\ref{fig:band}, where the difference between the upper and the lower value is  
approximately $10\%$. Thus, the general conclusions we can draw with respect to the properties of the unquenched propagator are quite insensitive to the extrapolation point used (and therefore, ultimately, to the value of $\lambda$). In what follows we will further explore the properties of $\Delta_{\s Q}(q^2)$ extrapolated towards the IR starting from  \mbox{$q^2 = 0.05\,\mbox{GeV}^2$} .

%%%%%%%%%%%%%%%%%%%%%%%%%%%%%%%%%%%%%%%%%%%%%%%%%%%%%%%%%%%%%%%%%%%%%%%%%
%             Fig.12  gluon propagator and gluon dressing
%%%%%%%%%%%%%%%%%%%%%%%%%%%%%%%%%%%%%%%%%%%%%%%%%%%%%%%%%%%%%%%%%%%%%%%%%%%%
\begin{figure}[!t]
%\hspace{-1.5cm}
\begin{minipage}[b]{0.45\linewidth}
\centering
\includegraphics[scale=0.5]{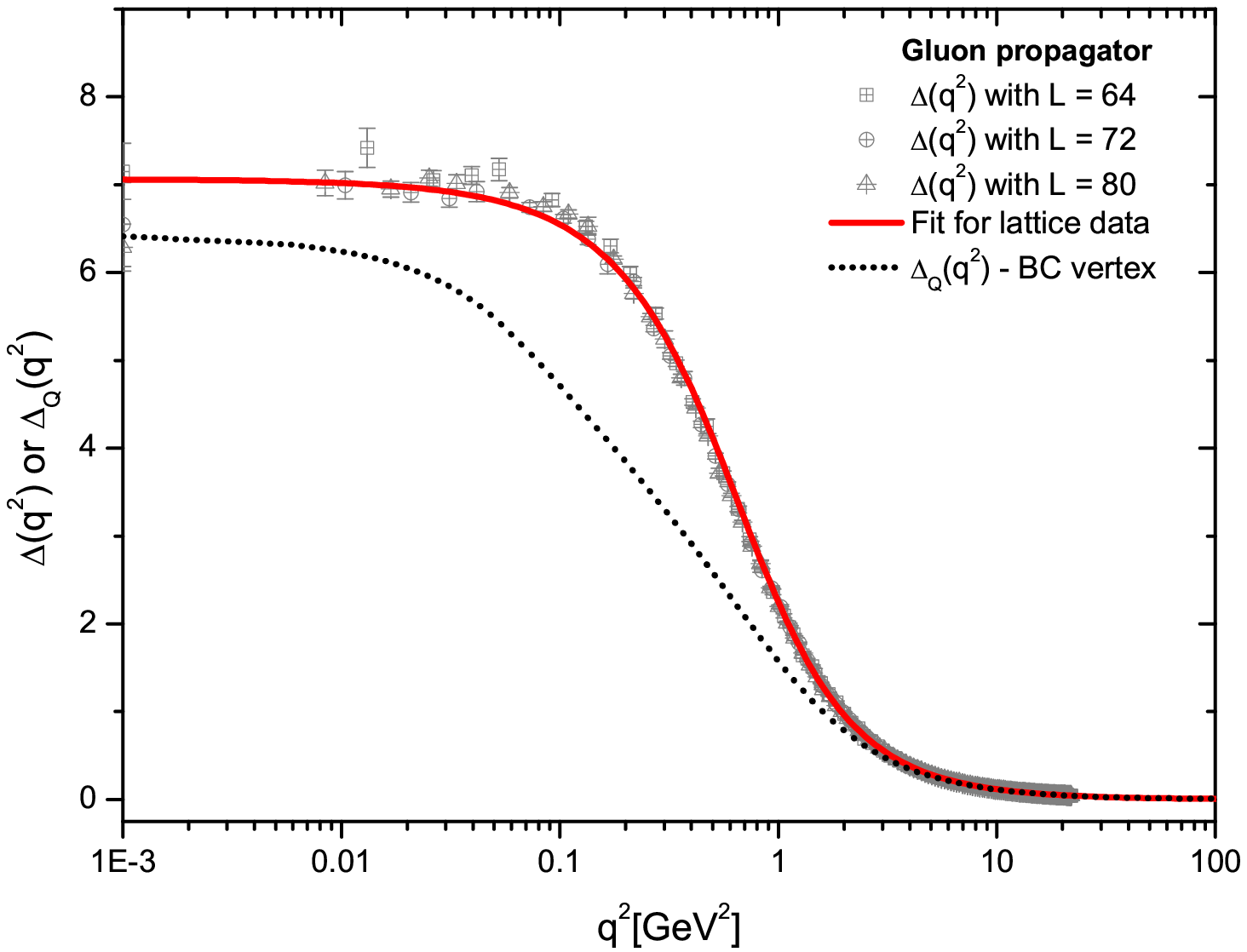} 
\end{minipage}
\hspace{0.5cm}
\begin{minipage}[b]{0.50\linewidth}
\hspace{-1.5cm}
\includegraphics[scale=0.5]{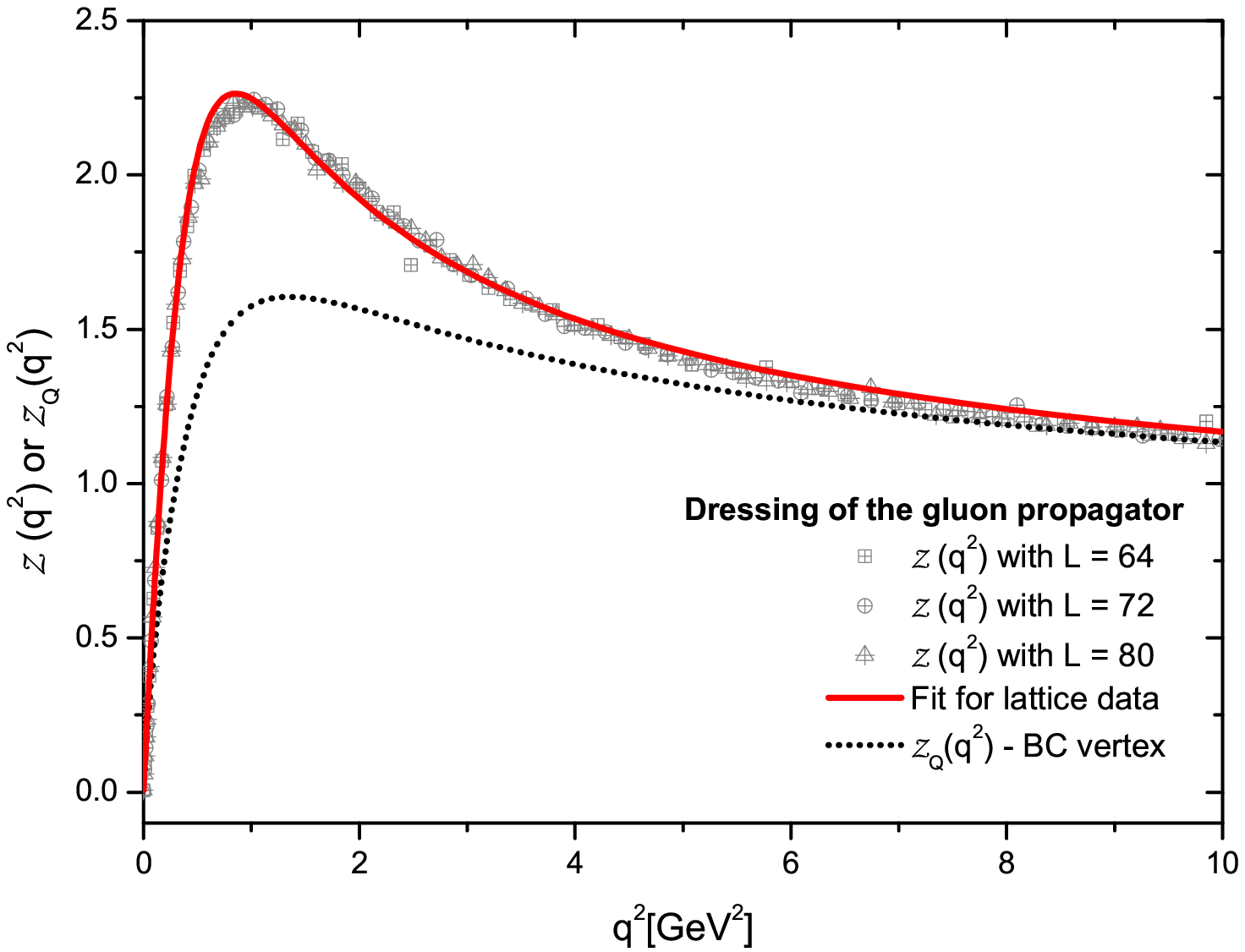}
\end{minipage}
\vspace{-0.75cm}
\caption{\label{fig:resultp} The quenched (solid red curve) and unquenched (dotted black curve) gluon propagators (left panel) and dressing functions (right panel). The unquenched case corresponds to the case where
the extrapolation starts at $q^2=0.05\,\mbox{GeV}^2$.} 
\end{figure}
%%%%%%%%%%%%%%%%%%%%%%%%%%%%%%%%%%%%%%%%%%%%%%%%%%%%%%%%%%%%%%%%%%%%

On the left panel of Fig.~\ref{fig:resultp}  
we superimpose the quenched lattice result of~\cite{Bogolubsky:2007ud} (solid red  curve)
and the unquenched result obtained from our calculation  (dotted  black
curve), while on the right panel we show a comparison of the corresponding dressing functions.
In the latter case notice that,
as expected,  both the  quenched and unquenched  curves vanish  at zero
momentum  transfer, and  their differences  in the  deep IR  region is
completely washed out. A direct comparison between the unquenched dressing function
computed here and that obtained in the unquenched lattice simulation of~\cite{Bowman:2007du} is postponed for the 
next subsection.

\n{v} The dependence of the unquenched solution
on the number of the flavors $n_f$ is next shown in \fig{fig:flavor}. As in previous plots, we show 
the quenched lattice data (solid red curve) as a benchmark, 
while different dashed and/or dotted curves correspond to different values of flavors: $n_f=1$ (dash-dotted green curve), $n_f=2$ (dotted black curve), and, finally, $n_f=3$ (dashed blue curve). Evidently, increasing the number of flavors results in a more suppressed gluon propagator. 
As can be seen clearly in \fig{fig:flavor},  
in the IR and intermediate regions 
the curves with more active quarks lie below the  
ones with fewer. This fact 
does not contradict the one-loop perturbative
behavior, given by \1eq{pert_unq}, stating that, in the UV,  
$\Delta_{\s Q}(q^2)$ increases for a higher number of quark families. Indeed,  we have 
checked that the perturbative behavior of $\Delta_{\s Q}(q^2)$ is recovered, due to 
a crossing that takes place around the renormalization point $\mu$, which makes the
curve for $n_f=3$ (dashed blue curve) go above all the others in the perturbative regime.

\n{vi} In Fig.~\ref{fig:1loop} we show another interesting property of $\Delta_{\s Q}(q^2)$. 
The dotted black curve represents  $\Delta_{\s Q}(q^2)$ obtained with the
nonperturbative  expression  for $\quark(q^2)$ given by \2eqs{quark-loop-full}{T123}; 
the  dash-dotted blue curve refers instead to the result of a simple one-loop calculation with a constant quark mass 
(see  \1eq{per2} in the Appendix). 
Notice that the latter result can be obtained by substituting  \mbox{$A_a=A_b=1$} and \mbox{$B_a=B_b=292$ MeV} into \2eqs{quark-loop-full}{T123}.
The difference between the two curves is at the few percent level, in agreement with the observation made before that the terms $T_2$ and $T_3$ are numerically subdominant (at the one-loop level these terms vanish, since $L_2=L_3=0$). 
These observations suggest that 
the nonperturbative quark loop diagram $(a_{11})$ appears to be rather insensitive to the running of the 
dynamical quark mass.

%%%%%%%%%%%%%%%%%%%%%%%%%%%%%%%%%%%%%%%%%%%%%%%%%%%%%%%%%%%%
%         Fig. 13
%%%%%%%%%%%%%%%%%%%%%%%%%%%%%%%%%%%%%%%%%%%%%%%%%%%%%%%%%%%%%%%%%%%
\begin{center}
\begin{figure}[!t]
\includegraphics[scale=0.5]{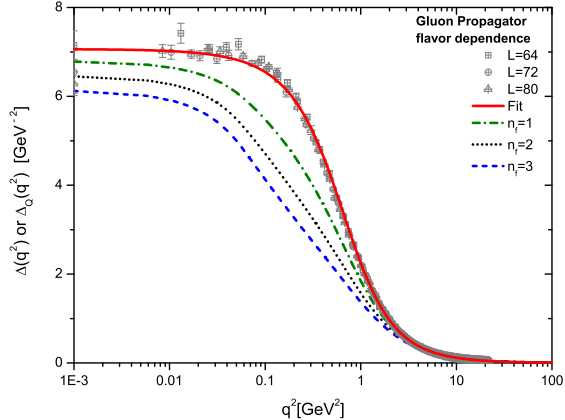}
\caption{\label{fig:flavor}The unquenched gluon propagator for different number of flavors: $n_f=1$ (dash-dotted green curve), $n_f=2$ (dotted black curve) and $n_f=3$  (dashed blue curve).}
\end{figure}
\end{center}
%%%%%%%%%%%%%%%%%%%%%%%%%%%%%%%%%%%%%%%%%%%%%%%%%%%%%%%%%%%%%%%%%%% 
\vspace{0.5cm}

%%%%%%%%%%%%%%%%%%%%%%%%%%%%%%%%%%%%%%%%%%%%%%%%%%%%%%%%%%%%
%        Fig 14 - one loop and comparison with heavier mass
%%%%%%%%%%%%%%%%%%%%%%%%%%%%%%%%%%%%%%%%%%%%%%%%%%%%%%%%%%%%%%%%%%%
\begin{center}
\begin{figure}[!t]
%\hspace{-1.5cm}
\begin{minipage}[b]{0.45\linewidth}
\centering
\includegraphics[scale=0.5]{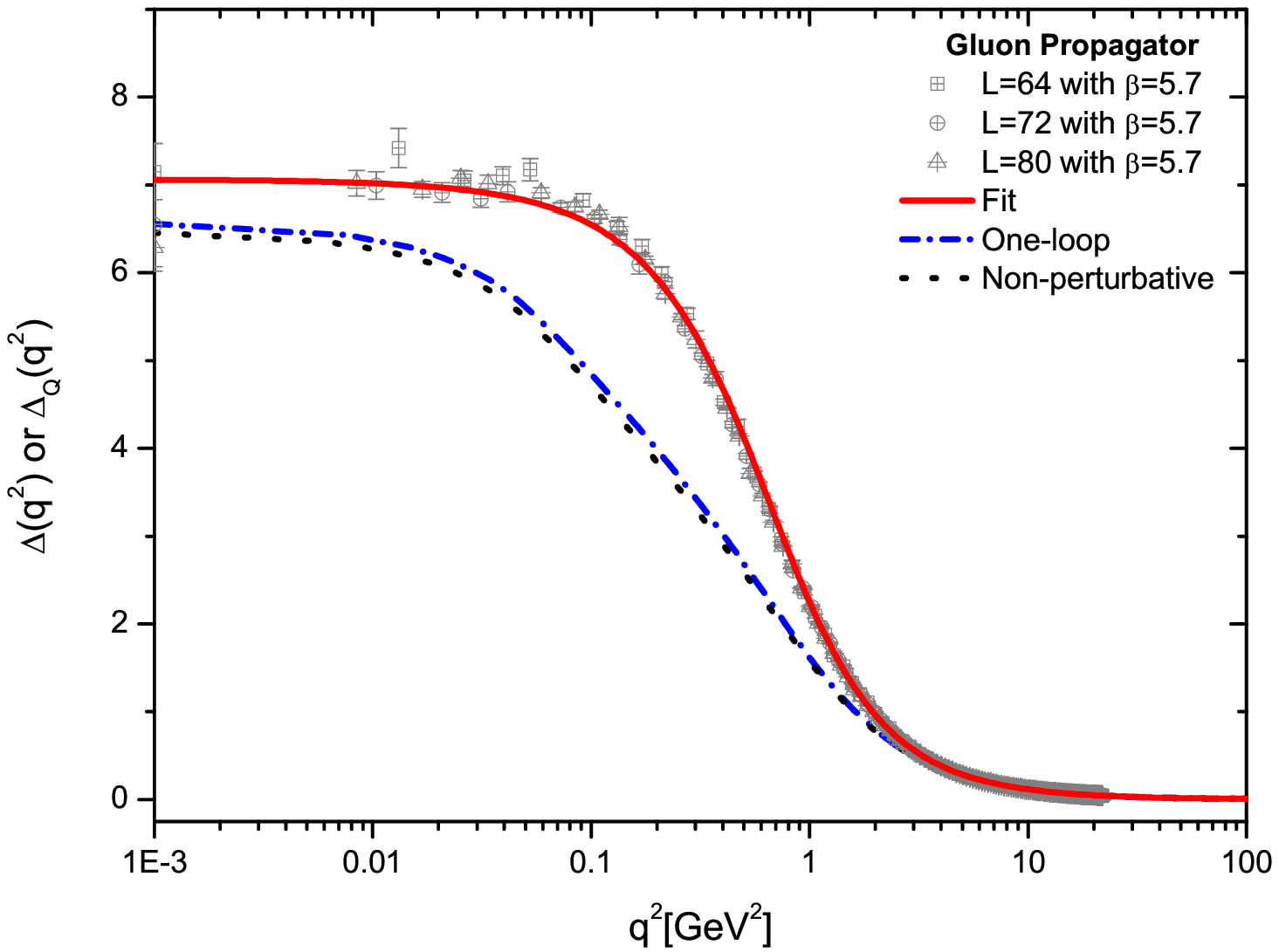} 
\end{minipage}
\hspace{0.5cm}
\begin{minipage}[b]{0.50\linewidth}
\hspace{-1.5cm}
\includegraphics[scale=0.5]{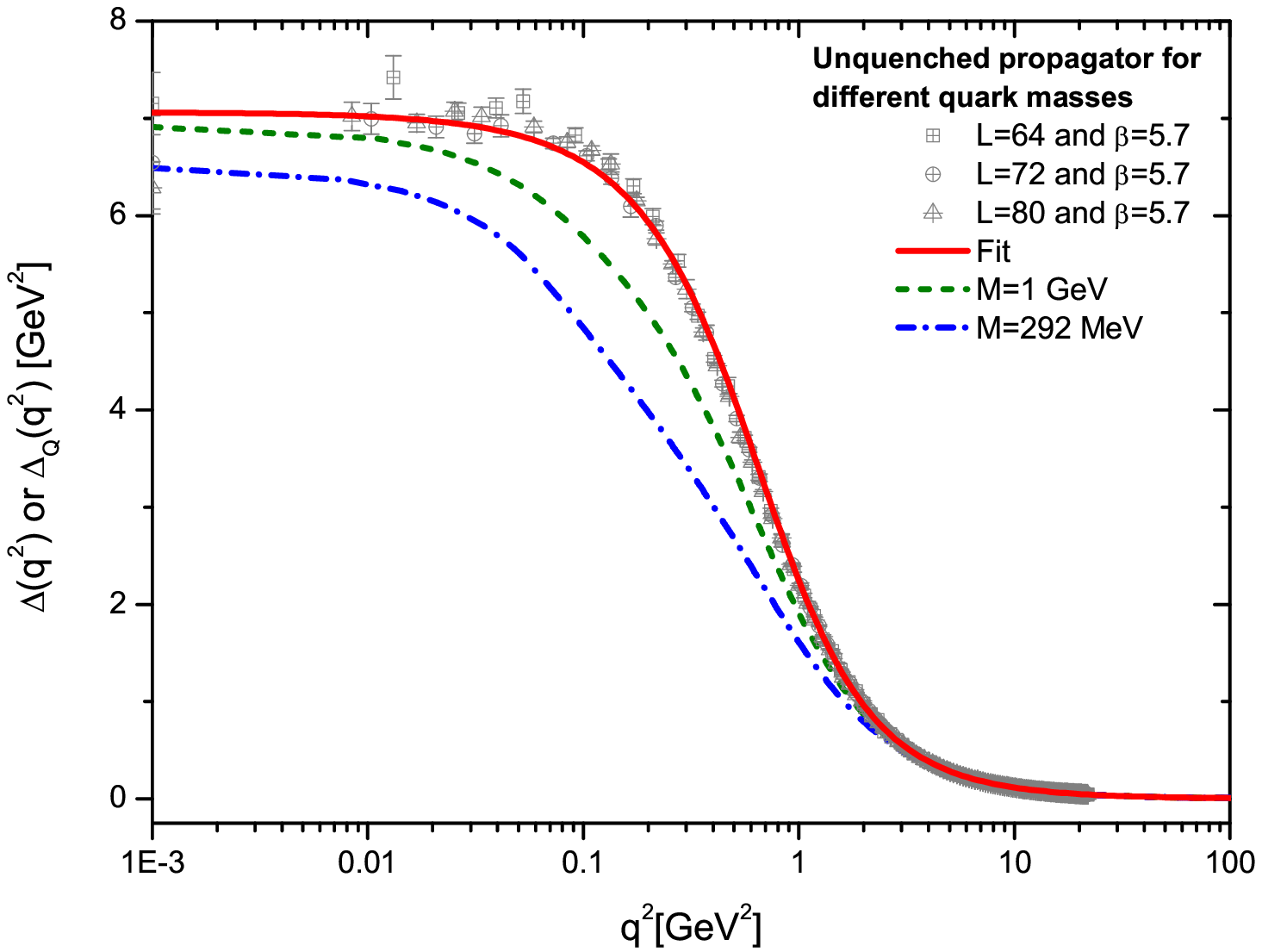}
\end{minipage}
\vspace{-0.75cm}
\caption{\label{fig:1loop} Left panel: The unquenched gluon propagator $\Delta_{\s Q}(q^2)$ for $n_f=2$ 
(dotted black curve) compared to the one-loop dressed result with 
a constant quark mass $M= 292\,$ MeV (dash-dotted blue curve). Right panel: The unquenched gluon propagator $\Delta_{\s Q}(q^2)$ for  $n_f=2$   with different  values of constant quark masses:  \mbox{$M=1$ GeV}(dashed green  curve) and \mbox{$M=292$ MeV} (dash-dotted blue curve).}
\end{figure}
\end{center}
%%%%%%%%%%%%%%%%%%%%%%%%%%%%%%%%%%%%%%%%%%%%%%%%%%%%%%%%%%%%%%%%%%% 

\n{vii} To check the decoupling of the heavier flavors, we also compare in the 
right panel of the same figure the result of the one-loop calculation with~$M=292$ MeV (dash-dotted blue curve) 
and \mbox{$M=1$ GeV} (dashed green curve). As we see clearly, 
the effect of the dynamical fermions on the gluon propagator becomes progressively suppressed as the quark mass increases.

\n{viii} Up to now, we have computed  the unquenched gluon propagator $\Delta_{\s Q}(q^2)$  for a particular fixed value
of the renormalization point $\mu$, namely \mbox{$\mu=4.3$ GeV}. It is well-known 
that, both  quenched and  unquenched  gluon propagators are $\mu$-dependent quantities, and therefore,
different choices of $\mu$ will lead to different results.

%%%%%%%%%%%%%%%%%%%%%%%%%%%%%%%%%%%%%%%%%%%%%%%%%%%%%%%%%%%%%%%%%%%%%%%%%%
%             Fig. 15 - mu dependence
%%%%%%%%%%%%%%%%%%%%%%%%%%%%%%%%%%%%%%%%%%%%%%%%%%%%%%%%%%%%%%%%%%%%%%%%%%%%
\begin{figure}[!t]
%\hspace{-1.5cm}
\begin{minipage}[b]{0.45\linewidth}
\centering
\includegraphics[scale=0.58]{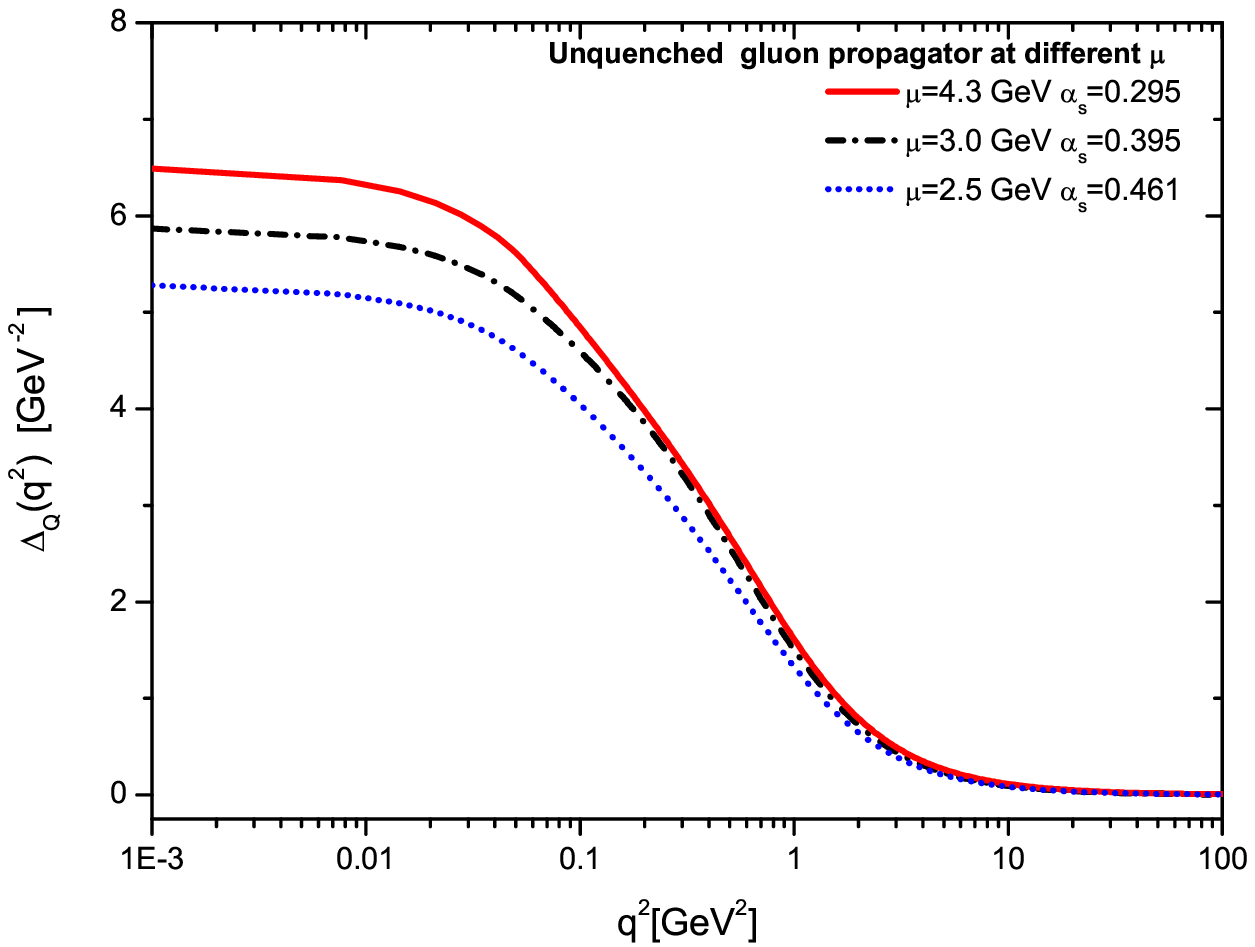}
\end{minipage}
\hspace{0.5cm}
\begin{minipage}[b]{0.45\linewidth}
\hspace{-1.5cm}
\includegraphics[scale=0.5]{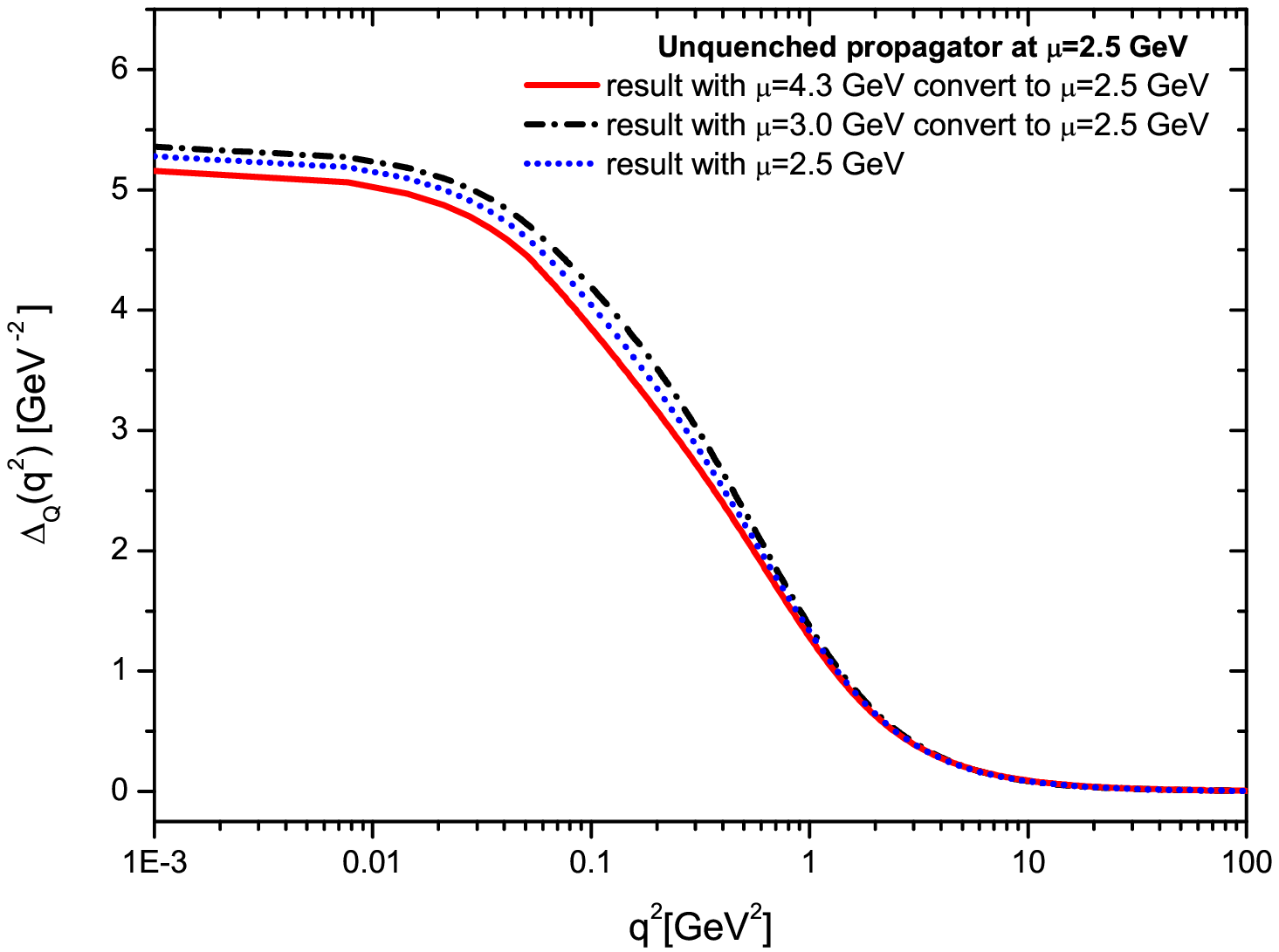}
\end{minipage}
\vspace{-0.5cm}
\caption{\label{fig:mu} Left panel: The $n_f=2$ unquenched gluon propagator renormalized at different values of $\mu$ and  $\alpha_s$:
 $\mu=4.3$ GeV and $\alpha_s=0.295$ (solid red curve), \mbox{$\mu=3.0$ GeV}  and \mbox{$\alpha_s=0.395$} (dash-dotted black curve), \mbox{$\mu=2.5$ GeV}  and \mbox{$\alpha_s=0.461$} (dotted blue curve).
Right panel: All curves showed on the left panel renormalized at the same point \mbox{$\mu=2.5$ GeV} using the Eq.~(\ref{ren_g2}).} 
\end{figure}
%%%%%%%%%%%%%%%%%%%%%%%%%%%%%%%%%%%%%%%%%%%%%%%%%%%%%%%%%%%%%%%%%%%%   

In order to address quantitatively this effect, 
on the left panel of Fig.~\ref{fig:mu} we show  $\Delta_{\s Q}(q^2)$ 
with $n_f=2$ for three different values of $\mu$: (i) \mbox{$\mu=2.5$ GeV} and {$\alpha_s=0.461$} (dotted blue curve); 
(ii) \mbox{$\mu=3.0$ GeV} and {$\alpha_s=0.395$} (dash-dotted black curve), and 
(iii) \mbox{$\mu=4.3$ GeV} and {$\alpha_s=0.295$} (solid red curve). Details on how
the values of $\alpha_s$ corresponding to each renormalization point were determined can be found in~\cite{Aguilar:2009nf, Aguilar:2011yb}.   
From \fig{fig:mu} one can then clearly see that higher values of $\mu$ correspond
to larger  values of  $\Delta_{\s Q}(0)$, which is basically the same pattern observed for the quenched case.

\n{ix} 
Finally, one important property that relates the gluon propagators renormalized at different values of $\mu$ is 
the  multiplicative renormalizability, which allows one to connect a set of  data renormalized 
at $\mu$ with a corresponding set renormalized at $\nu$, through the relation
\be
\Delta_{\s Q}(q^2,\mu^2)=\frac{\Delta_{\s Q}(q^2,\nu^2)}{\mu^2\Delta_{\s Q}(\mu^2,\nu^2)}\,.
\label{ren_g2}
\ee
On the right panel of Fig.~\ref{fig:mu} we check  how
$\Delta_{\s Q}(q^2)$ behaves under changes of $\mu$ using \1eq{ren_g2}. 
Evidently, multiplicative renormalizability would require that the three curves lie on top of each other; 
however we see that there is a minor difference between them (at the $4\%$ level),  
whose origin might be related to the fact 
that, as discussed in the previous section, in our computation the renormalization procedure 
was carried out subtractively instead of multiplicatively.

\subsection{\label{latt}Comparison with the lattice data}

In this final subsection   
we carry our a comparison between the results
we found for the gluon dressing function, $\dress_\s{Q}(q^2)$,  
and the data obtained from the unquenched lattice simulation of Ref.~\cite{Bowman:2007du}. 
We remind the reader that, according to the convention introduced below \1eq{qprop}, 
we will denote by $M$ (with the appropriate flavor index)
the value of the corresponding running quark mass $\qm(p^2)$ at $p^2=0$.

On the left panel of  Fig.~\ref{fig:lattice}, we show 
the $2+1$ flavor QCD lattice data  renormalized at 
\mbox{$\mu=4.3$ GeV}(gray open circles), together 
with the results obtained applying our calculational procedure, with two 
light quarks of mass \mbox{$M_{u/d}=292$ MeV} and one 
heavier of \mbox{$M_{h}=500$ MeV} (solid red curve). 
It is important to mention that the above ranges of quark masses are consistent with 
the values generally employed
in phenomenological calculations~\cite{Maris:1997tm, ElBennich:2011py}.

We clearly see that the overall shape of the calculated curves display a nice agreement with the data 
in a sizable range of momenta. The region where the difference
between the curves is more pronounced is around  \mbox{$q=1.25$ GeV}, exactly where
the peak of $\dress_\s{Q}(q^2)$ is located. However, 
observe that, even in this least favorable region, the difference between these curves is no greater than $10\%$.

%%%%%%%%%%%%%%%%%%%%%%%%%%%%%%%%%%%%%%%%%%%%%%%%%%%%%%%%%%%%%%%%%%%%%%%%%
%             Fig.16  lattice comparision - gluon dressing function
%%%%%%%%%%%%%%%%%%%%%%%%%%%%%%%%%%%%%%%%%%%%%%%%%%%%%%%%%%%%%%%%%%%%%%%%%%%%
\begin{figure}[!t]
%\hspace{-1.5cm}
\begin{minipage}[b]{0.45\linewidth}
\centering
\includegraphics[scale=0.5]{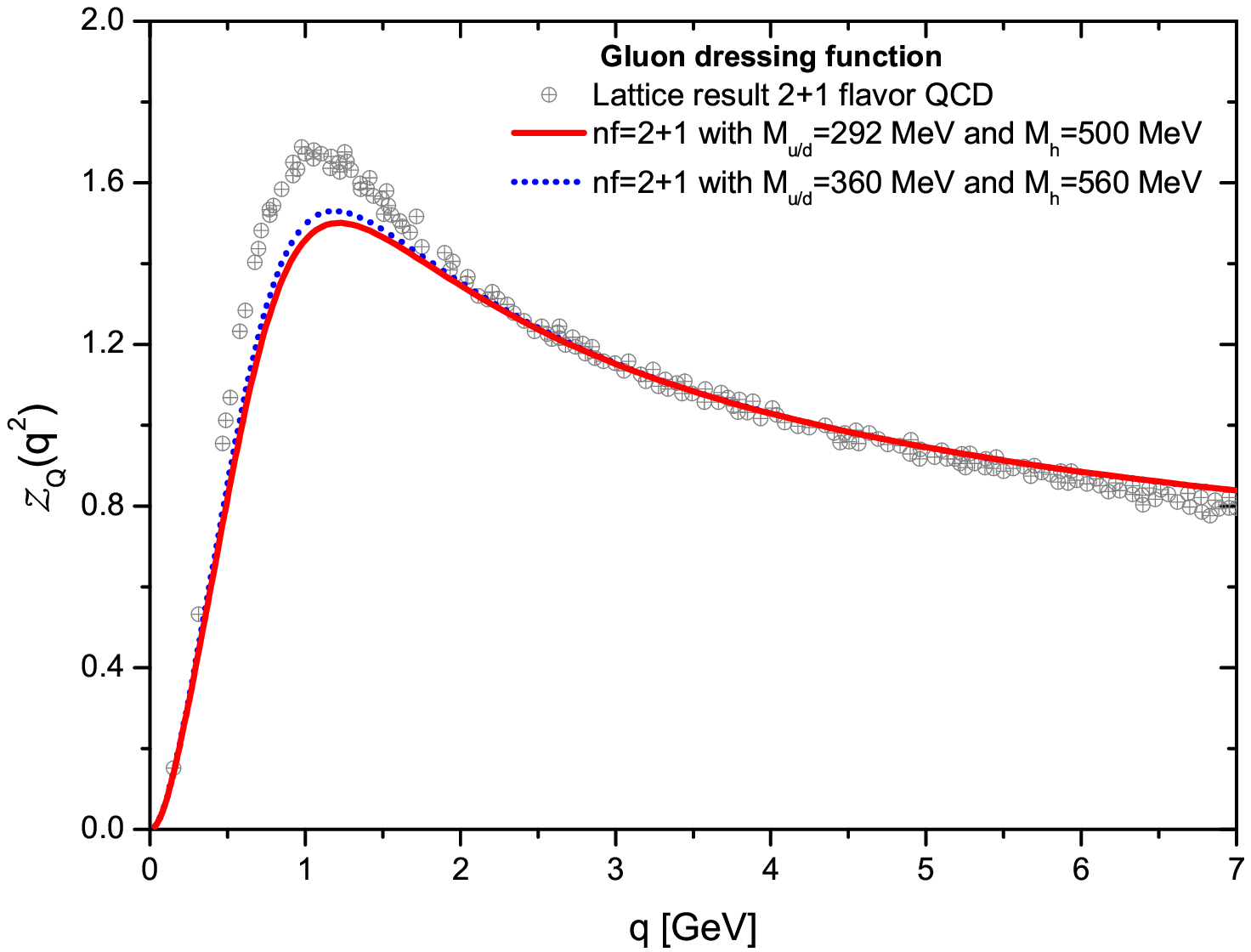} 
\end{minipage}
\hspace{0.5cm}
\begin{minipage}[b]{0.50\linewidth}
\hspace{-1.5cm}
\includegraphics[scale=0.5]{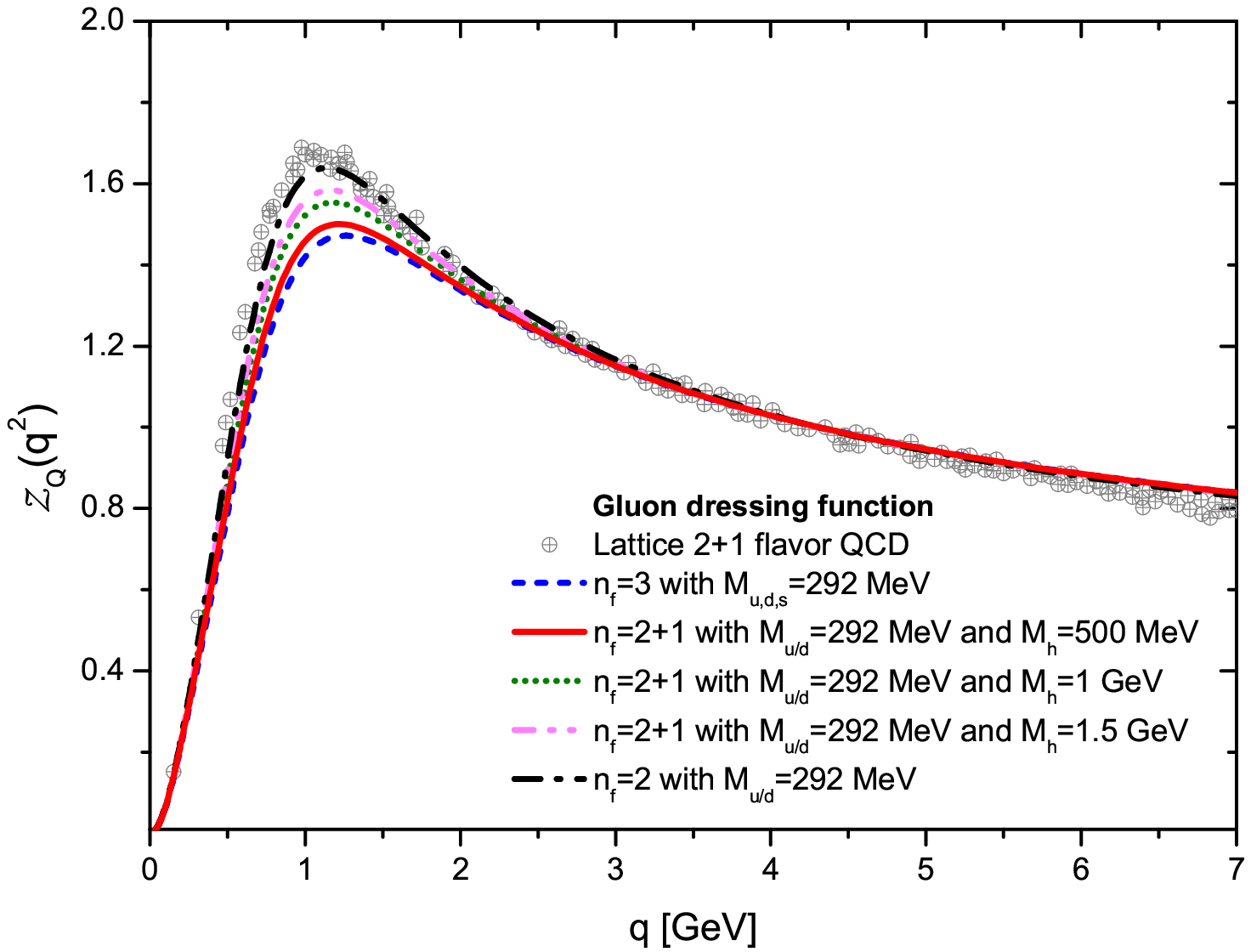}
\end{minipage}
\vspace{-0.75cm}
\caption{\label{fig:lattice} Left panel: The unquenched gluon dressing function $\dress_\s{Q}(q^2)$
obtained in the Ref.~\cite{Bowman:2007du} (open gray circles) together with the one-loop result for two light quarks with \mbox{$M_{u/d}=292$ MeV}, and one heavier with \mbox{$M_{s}=500$ MeV} (continuous red curve) and the 
case where \mbox{$M_{u/d}=360$ MeV} and  \mbox{$M_{s}=560$ MeV} (dotted blue curve).  Right panel: The  one-loop $\dress_\s{Q}(q^2)$ for $n_f=2+1$ flavors. The  light quarks  have constant masses of \mbox{$M_{u/d}=292$ MeV} 
while for the heavier quark:  \mbox{$M_h=292$ GeV}(dashed blue curve), \mbox{$M_h=500$ MeV} (solid red curve), \mbox{$M_h=1.0$ GeV} (dotted green curve) and \mbox{$M_h=1.5$ GeV} (dashed with two dots magenta curve).} 
\end{figure}
%%%%%%%%%%%%%%%%%%%%%%%%%%%%%%%%%%%%%%%%%%%%%%%%%%%%%%%%%%%%%%%%%%%%

In addition, notice that on the same plot we display also the case 
where \mbox{$M_{u/d}=360$ MeV} and  \mbox{$M_{h}=560$ MeV} (dashed blue curve).
The way this latter set of mass values (and corresponding quark propagators) 
are obtained is by solving the quark gap equation using 
suitable values for the current masses.  
More specifically, for the light 
quarks (up/down) we use a current mass of \mbox{$14$ MeV}, while for the heavier one (strange) we use \mbox{$68$ MeV}, 
in agreement with the values quoted in  Ref.~\cite{Bowman:2007du}. Observe that the  dotted  blue curve indicates 
that the increase of masses produces a slight change in the peak of the dressing function.

It would be instructive to analyze how different choices for $M_h$
modify the form $\dress_\s{Q}(q^2)$ shown on the left panel of  Fig.~\ref{fig:lattice}.
As we have already shown on the right panel of Fig.~\ref{fig:1loop}, 
the gluon propagator becomes progressively suppressed as the quark mass increases. It is
therefore natural to expect that the gluon dressing function 
will be also affected by different choices of masses.

On the right panel of the Fig.~\ref{fig:lattice}, we show how $\dress_\s{Q}(q^2)$ varies with $M_h$.
In all curves the light quarks  have constant masses of \mbox{$M_{u/d}=292$ MeV},
whereas for the heavier quark we use:  \mbox{$M_h=292$ GeV} (dashed blue curve), \mbox{$M_h=500$ MeV} 
(solid red curve), \mbox{$M_h=1.0$ GeV} (dotted green curve) and \mbox{$M_h=1.5$ GeV} (dashed with two dots magenta curve).

As we clearly see, the peak of $\dress_\s{Q}(q^2)$ becomes more pronounced as we increase the value of $M_h$. 
Notice that the case where \mbox{$M_h=1.5$ GeV} (dashed with two dots magenta curve) is much closer to the lattice
data (open gray circles). 
In addition, if we keep increasing the heavy quark mass gradually, the observed trend of the results is  
to move progressively closer to the case where only two quarks are active (dash-dotted black curve), 
thus confirming the notion of decoupling of heavy flavors. To avoid any possible confusion caused by the striking proximity of 
the $n_f=2$ curve to the lattice data, we reiterate that the real comparison 
between the $(2+1)$ data and the corresponding $(2+1)$ SDE result is given on the left panel.
Evidently, the analogous comparison of the 
$n_f=2$ curve would require a corresponding set of lattice data, not available to us at present.

\section{\label{conc}Conclusions}

In this article we have presented a general method for estimating the effects 
that the ``unquenching'' induces on the (IR finite) gluon propagator, in the 
Landau gauge. The basic assumption of the method followed has been that the 
main bulk of the effect originates from the ``one-loop dressed'' quark diagram, while the 
rest of the contributions is considered to be subleading. We have restricted the 
applicability of this approach to a small number of quark families ($n=1,2,3$), where we assume 
the presence  of the quarks does not alter 
qualitatively the behavior of the quenched propagator. In particular, we expect that the 
crucial property of IR finiteness will persist, \ie the gluon mass 
generating mechanism will not be distorted by the inclusion of a few quark families. 
In fact, throughout our analysis we use the quenched gluon propagator obtained in $SU(3)$ lattice 
simulations as our point of reference, and estimate the deviations induced to it 
by the quarks.

The nonperturbative calculation of the quark loop proceeds by means of two suitable
Ans\"atze for the fully dressed quark-gluon vertex $\widehat{\Gamma}_{\mu}$, enforcing the exact transversality of the 
resulting contribution.
The use of the PT-BFM formalism simplifies 
the form of these Ans\"atze
considerably, due to the  
 ``abelianization'' that it induces, given that 
the corresponding Green's functions, when contracted with respect to the momentum
carried by the background leg, satisfy linear ghost free WIs instead of the usual nonlinear STIs. 
This fact, in turn, avoids the 
explicit reference to the quark-ghost kernel, 
which appears in the standard STI satisfied by the conventional 
quark-gluon vertex $\Gamma_{\mu}$ [the $H$ auxiliary function of \1eq{STI}]. Of course, one cannot completely eliminate any dependence on $H$, 
for the simple reason that it affects the quark gap equation that determines 
the quantities $A(p)$ and $B(p)$, namely the nonperturbative Dirac components of the quark propagator;
this happens because, as explained in~\cite{Aguilar:2010cn},   
the quark-gluon vertex entering in the gap  equation is $\Gamma_{\mu}$ and not $\widehat{\Gamma}_{\mu}$.
Given that the structure of the  quark-ghost kernel is largely unexplored 
(for an SD estimate of one of its form-factors, see~\cite{Aguilar:2010cn}), 
reducing the dependence of the answer on it is clearly advantageous. 

The main results of our study is that 
the inclusion of the quark loop(s) induces a suppression in the intermediate and IR momentum regions, 
with respect to the quenched case.  As emphasized in the main text, the actual saturation point 
of the unquenched propagator, i.e., the value $\Delta_{\s Q}(0)$, normally associated with the IR value of the 
dynamical gluon mass, $m^2(0)$, is not possible to determine at present, 
despite the fact that the quark-loop contribution to the corresponding gluon self-energy vanishes 
at $q^2=0$, by virtue of a powerful identity.
The reason is that the momentum evolution of the gluon mass   
depends (in a yet not fully determined way) on the structure of the gluon propagator through the {\it entire range} of physical momenta; thus, 
the suppression of the propagator due to the inclusion of the quarks is expected to modify the value of $m^2(0)$. 
In this work we have adopted a simple  
hand-waving approach for estimating $\Delta_{\s Q}(0)$. Specifically, given that the unquenched propagator in the IR and intermediate regions 
is consistently below the corresponding quenched curve, we have simply extrapolated towards the point $q^2=0$. 
In practice, the outcome of this simple procedure depends to some extent on the extrapolation details 
(in particular, what one considers as the last ``faithful'' point), and therefore one can only   
determine a certain range of ``reasonable'' values for $\Delta_{\s Q}(0)$. 

The uncertainty associated with the 
determination of the saturation point is practically eradicated if one considers 
instead of the gluon propagator its corresponding dressing function. 
This latter quantity, when compared to the corresponding dressing function of the quenched lattice propagator,  
clearly demonstrates the aforementioned suppression in the IR and intermediate regions 
induced by the inclusion of the quarks.  
The unquenched dressing function obtained through our procedure appears to be in rather good agreement 
with the lattice results available in the literature.

There are certain theoretical improvements, which, if successfully implemented,  
would put the proposed approach on a more solid ground.
To begin with, it is clear that the full SDE treatment of the problem at hand would entail 
the {\it simultaneous} treatment of a complicated set of coupled integral equations, in the spirit 
presented in~\cite{Fischer:2003rp,Fischer:2005en}, in the context of the scaling solutions. 
This type of global treatment appears to be beyond our present 
calculational powers, mainly due to the plethora of additional technical complications intrinsic to the massive solutions.
Instead, we have adopted a step-by-step procedure;
for example, the quark-gap equation has been solved ``in isolation'', and the obtained solutions have been fed into 
the equations determining the quark-loop, and so on. To be sure, this latter procedure might interfere with 
the nonlinear propagation of certain effects, leading to the corresponding  
amplification or suppression of various features, and may require additional refinements. 

The renormalization properties of the relevant integral equations
constitute a commonly known 
source of theoretical uncertainty, due to the mishandling of the overlapping divergences induced by the 
well-known intrinsic ambiguity of the gauge-technique, related to the unspecified transverse 
(automatically conserved) part of the vertices. In particular, the BC and CP expressions employed here 
for the quark-gluon vertex do not fully respect the property of multiplicative renormalizability, which, in turn, 
leads to dependences on the renormalization point that are not always in accordance with those 
dictated by the renormalization group. The propagation of such discrepancies to our predictions has 
been studied numerically, and appears to be relatively suppressed. However, more work is clearly needed 
in order to eliminate the spurious $\mu$-dependences. In this vain, it would be interesting, 
albeit logistically cumbersome, 
to explore the effects that other forms of the quark-gluon vertex 
might have on our predictions, such as those reported  in~\cite{Kizilersu:2009kg,Bashir:2011dp}.

Finally, the reliable calculation of the saturation point $\Delta_{\s Q}(0)$ mentioned above 
hinges explicitly on the derivation of a fully self-consistent integral equation, that would determine the 
momentum evolution of the dynamical gluon mass, both in the quenched case and in the presence of quarks. 
The derivation of such a complete equation is conceptually and technically rather non-trivial, and 
is the subject of an ongoing investigation, whose results will be hopefully presented soon. 

\acknowledgments 

The research of J.~P. is supported by the Spanish MEYC under 
grant FPA2011-23596. The work of  A.C.A  is supported by the Brazilian
Funding Agency CNPq under the grant 305850/2009-1 and project 474826/2010-4 .

\appendix
\section{The perturbative One-loop case}

The text-book perturbative calculation of diagram $a_{11}$ yields (with $d_f =1/2$) 
\be
\quark^{[1]}(q^2) =- \frac{2g^2}{d-1}\kint\,\frac{dM^2-(d-2)(k^2+k\cdot q)}{(k^2-M^2)[(k+q)^2-M^2]}\,.
\label{per1}
\ee
where $M$ denotes a constant (momentum-independent) mass. 
Note that the ``hat'' in this case is redundant, because, at one loop, the conventional and BFM results 
coincide. The result of \1eq{per1} may be directly recovered from the general case presented in section \ref{nonpertloops},
by setting $\qm(p)=M$, $A(p)=1$, $L_1=1$, $L_2=L_3=0$
in Eqs.(\ref{quark-loop-full}) and (\ref{T123}).

It is elementary to establish that 
\be
\quark^{[1]}(0) =0 \,,
\label{per1mass}
\ee
by virtue of the basic identity
\be
\int_k \frac{k^2}{(k^2-M^2)^2} = \frac{d}{2} \int_k \frac{1}{k^2-M^2} \,,
\label{id1}
\ee
or, equivalently, 
\be
2 M^2\int_k \frac{1}{(k^2-M^2)^2} = (d-2)\int_k \frac{1}{k^2-M^2} \,, 
\label{id2}
\ee
whose validity may be easily verified  following the 
integration rules of dimensional regularization. These exact same identities appear 
in the standard one-loop 
calculation of the photon vacuum polarization, both in normal QED and  
in scalar QED, and enforces the masslessness of the photon~\cite{Aguilar:2009ke}.

The property of (\ref{per1mass}) becomes manifest through the use of (\ref{id2}), which 
allows one to cast (\ref{per1}) into the form 
\be
\quark^{[1]}(q^2) = - \frac{g^2}{d-1}\left\{(d-2) q^2 I(q^2) + 4 M^2 \left[I(q^2)-I(0)\right]\right\},
\label{per2}
\ee
where
\be
I(q^2) = \kint \frac{1}{(k^2-M^2)[(k+q)^2-M^2]}\,,
\ee
or, equivalently, defining 
\be
u^2(q^2) \equiv q^2 x(x-1) + M^2,
\ee
we have 
\be
\quark^{[1]}(q^2)=-\frac{g^2}{d-1}\left\{ (d-2) q^2 I(q^2) -i\,\frac{M^2}{4\pi^2}\int_0^1\!\diff{}x\ln \frac{u^2(q^2)}{M^2}
\right\}.
\ee
Finally, the renormalized expression for $\quark^{(1)}(q^2)$ in the MOM scheme is given by 
\be
\quarkren^{[1]}(q^2) = \quark^{[1]}(q^2) - \frac{q^2}{\mu^2} \quark^{[1]}(\mu^2) \,.
\ee
giving as a result
\be
\quarkren^{[1]}(q^2) = \frac{i\alpha_s}{6\pi} 
\left\{
q^2 \int_0^1\!\diff{}x\ln \frac{u^2(q^2)}{u^2(\mu^2)} 
+ 2 M^2 \left[\int_0^1\!\diff{}x\ln\frac{u^2(q^2)}{M^2} - \frac{q^2}{\mu^2}
\int_0^1\!\diff{}x\ln \frac{u^2(\mu^2)}{M^2}
\right] \right\}.
\ee
Evidently, for $q^2$ and $\mu^2$ much larger than $M^2$, one obtains the standard logarithmic correction
\be 
\quarkren^{[1]}(q^2) = \frac{i\alpha_s}{6\pi} q^2 \ln (-q^2/\mu^2).
\label{qr1}
\ee


\begin{thebibliography}{99}


%\cite{Cucchieri:2007md}
\bibitem{Cucchieri:2007md}
A.~Cucchieri and T.~Mendes,
%``What's up with IR gluon and ghost propagators in Landau gauge? A puzzling
%answer from huge lattices,''
PoS {\bf LAT2007}, 297 (2007).
%  [arXiv:0710.0412 [hep-lat]].
%%CITATION = POSCI,LAT2007,297;%%

%\cite{Cucchieri:2007rg}
\bibitem{Cucchieri:2007rg}
A.~Cucchieri and T.~Mendes,
%``Constraints on the IR behavior of the gluon propagator in Yang-Mills
%theories,''
Phys.\ Rev.\ Lett.\  {\bf 100}, 241601 (2008).
%  [arXiv:0712.3517 [hep-lat]].
%%CITATION = PRLTA,100,241601;%%  

%\cite{Cucchieri:2009zt}
\bibitem{Cucchieri:2009zt}
A.~Cucchieri and T.~Mendes,
%``Landau-gauge propagators in Yang-Mills theories at beta = 0: massive
%solution versus conformal scaling,''
Phys.\ Rev.\  D {\bf 81}, 016005 (2010).
%[arXiv:0904.4033 [hep-lat]].
%%CITATION = PHRVA,D81,016005;%%

%\cite{Cucchieri:2011ga}
\bibitem{Cucchieri:2011ga}
 A.~Cucchieri and T.~Mendes,
%``Further Investigation of Massive Landau-Gauge Propagators in the Infrared
%Limit,''
PoS {\bf LATTICE2010}, 280 (2010).
%  [arXiv:1101.4537 [hep-lat]].
%%CITATION = POSCI,LATTICE2010,280;%%


%\cite{Kamleh:2007ud}
\bibitem{Kamleh:2007ud} 
  W.~Kamleh, P.~O.~Bowman, D.~B.~Leinweber, A.~G.~Williams and J.~Zhang,
  %``Unquenching effects in the quark and gluon propagator,''
  Phys.\ Rev.\ D {\bf 76}, 094501 (2007).
%  [arXiv:0705.4129 [hep-lat]].
  %%CITATION = ARXIV:0705.4129;%%


%\cite{Bowman:2007du}
\bibitem{Bowman:2007du} 
  P.~O.~Bowman, U.~M.~Heller, D.~B.~Leinweber, M.~B.~Parappilly, A.~Sternbeck, L.~von Smekal, A.~G.~Williams and J.~-b.~Zhang,
  %``Scaling behavior and positivity violation of the gluon propagator in full QCD,''
  Phys.\ Rev.\ D {\bf 76}, 094505 (2007).
%  [hep-lat/0703022 [HEP-LAT]].
  %%CITATION = HEP-LAT/0703022;%%



%\cite{Bogolubsky:2007ud}
\bibitem{Bogolubsky:2007ud}
 I.~L.~Bogolubsky, E.~M.~Ilgenfritz, M.~Muller-Preussker and A.~Sternbeck,
  %``The Landau gauge gluon and ghost propagators in 4D SU(3) gluodynamics in
%large lattice volumes,''
PoS {LATTICE}, 290 (2007).
%  [arXiv:0710.1968 [hep-lat]].
%%CITATION = POSCI,LATTICE,290;%%

%\cite{Bogolubsky:2009dc}
\bibitem{Bogolubsky:2009dc}
 I.~L.~Bogolubsky, E.~M.~Ilgenfritz, M.~Muller-Preussker and A.~Sternbeck,
%``Lattice gluodynamics computation of Landau gauge Green's functions in the
%deep infrared,''
Phys.\ Lett.\  B {\bf 676}, 69 (2009).
% [arXiv:0901.0736 [hep-lat]].
%%CITATION = PHLTA,B676,69;%%

%\cite{Oliveira:2008uf}
\bibitem{Oliveira:2008uf}
  O.~Oliveira, P.~J.~Silva,
  %``Does The Lattice Zero Momentum Gluon Propagator for Pure Gauge SU(3) Yang-Mills Theory Vanish in the Infinite Volume Limit?,''
  Phys.\ Rev.\  {\bf D79}, 031501 (2009).
%  [arXiv:0809.0258 [hep-lat]].

%\cite{Oliveira:2009eh}
\bibitem{Oliveira:2009eh}
O.~Oliveira and P.~J.~Silva,
%``The lattice infrared Landau gauge gluon propagator: the infinite volume
%limit,''
PoS {\bf LAT2009}, 226 (2009).
% [arXiv:0910.2897 [hep-lat]].
%%CITATION = POSCI,LAT2009,226;%%

%%%%%%%%
%  SDE
%%%%%%%%


%\cite{Alkofer:2000wg}
\bibitem{Alkofer:2000wg}
 R.~Alkofer, L.~von Smekal,
 %``The Infrared behavior of QCD Green's functions: Confinement 
% dynamical symmetry breaking, and hadrons as relativistic bound states,''
 Phys.\ Rept.\  {\bf 353}, 281 (2001).
%  [hep-ph/0007355].


%\cite{Fischer:2006ub}
\bibitem{Fischer:2006ub}
 C.~S.~Fischer,
 %``Infrared properties of QCD from Dyson-Schwinger equations,''
 J.\ Phys.\ G {\bf G32}, R253-R291 (2006).
%  [hep-ph/0605173].

%\cite{Aguilar:2006gr}
\bibitem{Aguilar:2006gr}
A.~C.~Aguilar and J.~Papavassiliou,
%``Gluon mass generation in the PT-BFM scheme,''
JHEP {\bf 0612}, 012 (2006).
%  [arXiv:hep-ph/0610040].
%%CITATION = JHEPA,0612,012;%%



%\cite{Binosi:2007pi}
\bibitem{Binosi:2007pi}
D.~Binosi and J.~Papavassiliou,
%``Gauge-invariant truncation scheme for the Schwinger-Dyson equations of
%QCD,''
Phys.\ Rev.\  D {\bf 77}(R), 061702 (2008).
%%CITATION = PHRVA,D77,061702;%%


%\cite{Aguilar:2008xm}
\bibitem{Aguilar:2008xm}
  A.~C.~Aguilar, D.~Binosi and J.~Papavassiliou,
  %``Gluon and ghost propagators in the Landau gauge: Deriving lattice results
  %from Schwinger-Dyson equations,''
  Phys.\ Rev.\  D {\bf 78}, 025010 (2008).
  %[arXiv:0802.1870 [hep-ph]].
  %%CITATION = PHRVA,D78,025010;%%


%\cite{Binosi:2009qm}
\bibitem{Binosi:2009qm}  
D.~Binosi and J.~Papavassiliou,
%``Pinch Technique: Theory and Applications,''
Phys.\ Rept.\  {\bf 479}, 1-152 (2009).
%[arXiv:0909.2536 [hep-ph]].

  
%\cite{RodriguezQuintero:2011vw}
\bibitem{RodriguezQuintero:2011vw}
  J.~Rodriguez-Quintero,
  %``A brief comment on the similarities of the IR solutions for the ghost propagator DSE in Landau and Coulomb gauges,''
  Phys.\ Rev.\  {\bf D83}, 097501 (2011).
%  [arXiv:1103.0924 [hep-ph]].

%\cite{RodriguezQuintero:2010wy}
\bibitem{RodriguezQuintero:2010wy}
  J.~Rodriguez-Quintero,
  %``On the massive gluon propagator, the PT-BFM scheme and the low-momentum behaviour of decoupling and scaling DSE solutions,''
  JHEP {\bf 1101}, 105 (2011).
%  [arXiv:1005.4598 [hep-ph]].
   
 
%\cite{Boucaud:2010gr}
\bibitem{Boucaud:2010gr}
  Ph.~Boucaud, M.~E.~Gomez, J.~P.~Leroy, A.~Le Yaouanc, J.~Micheli, O.~Pene, J.~Rodriguez-Quintero,
  %``The low-momentum ghost dressing function and the gluon mass,''
  Phys.\ Rev.\  {\bf D82}, 054007 (2010).
%  [arXiv:1004.4135 [hep-ph]].


%\cite{Boucaud:2008gn}
\bibitem{Boucaud:2008gn}
  Ph.~Boucaud, F.~De Soto, J.~P.~Leroy, A.~Le Yaouanc, J.~Micheli, O.~Pene and J.~Rodriguez-Quintero,
  %``Ghost-gluon running coupling, power corrections and the determination of
  %Lambda(MS-bar),''
  Phys.\ Rev.\  D {\bf 79}, 014508 (2009).
 % [arXiv:0811.2059 [hep-ph]].
  %%CITATION = PHRVA,D79,014508;%%


%\cite{Boucaud:2008ji}
\bibitem{Boucaud:2008ji} 
  P.~Boucaud, J-P.~Leroy, A.~L.~Yaouanc, J.~Micheli, O.~Pene and J.~Rodriguez-Quintero,
  %``IR finiteness of the ghost dressing function from numerical resolution of the ghost SD equation,''
  JHEP {\bf 0806}, 012 (2008).
%  [arXiv:0801.2721 [hep-ph]].
  %%CITATION = ARXIV:0801.2721;%%



%\cite{Fischer:2008uz}
\bibitem{Fischer:2008uz} 
  C.~S.~Fischer, A.~Maas and J.~M.~Pawlowski,
  %``On the infrared behavior of Landau gauge Yang-Mills theory,''
  Annals Phys.\  {\bf 324}, 2408 (2009).
%  [arXiv:0810.1987 [hep-ph]].
  %%CITATION = ARXIV:0810.1987;%%



%\cite{Szczepaniak:2010fe}
\bibitem{Szczepaniak:2010fe}
A.~P.~Szczepaniak and H.~H.~Matevosyan,
%``A model for QCD ground state with magnetic disorder,''
Phys.\ Rev.\  D {\bf 81}, 094007 (2010).
%  [arXiv:1003.1901 [hep-ph]].
%%CITATION = PHRVA,D81,094007;%%

%\cite{Aguilar:2004sw}
\bibitem{Aguilar:2004sw} 
  A.~C.~Aguilar and A.~A.~Natale,
  %``A Dynamical gluon mass solution in a coupled system of the Schwinger-Dyson equations,''
  JHEP {\bf 0408}, 057 (2004).
%  [hep-ph/0408254].
  %%CITATION = HEP-PH/0408254;%%

%\cite{Dudal:2008sp}
\bibitem{Dudal:2008sp}
D.~Dudal, J.~A.~Gracey, S.~P.~Sorella, N.~Vandersickel and H.~Verschelde,
%``A refinement of the Gribov-Zwanziger approach in the Landau gauge: infrared
%propagators in harmony with the lattice results,''
Phys.\ Rev.\  D {\bf 78}, 065047 (2008).
%  [arXiv:0806.4348 [hep-th]].
%%CITATION = PHRVA,D78,065047;%%


%\cite{Dudal:2010tf}
\bibitem{Dudal:2010tf}
  D.~Dudal, O.~Oliveira, N.~Vandersickel,
  %``Indirect lattice evidence for the Refined Gribov-Zwanziger formalism and the gluon condensate $\braket{A^2}$ in the Landau gauge,''
  Phys.\ Rev.\  {\bf D81}, 074505 (2010).
%  [arXiv:1002.2374 [hep-lat]].


%\cite{Dudal:2011gd}
\bibitem{Dudal:2011gd} 
  D.~Dudal, S.~P.~Sorella and N.~Vandersickel,
  %``The dynamical origin of the refinement of the Gribov-Zwanziger theory,''
  Phys.\ Rev.\ D {\bf 84}, 065039 (2011).
%  [arXiv:1105.3371 [hep-th]].
  %%CITATION = ARXIV:1105.3371;%%


%\cite{Kondo:2011ab}
\bibitem{Kondo:2011ab} 
  K.~-I.~Kondo,
  %``A low-energy effective Yang-Mills theory for quark and gluon confinement,''
  Phys.\ Rev.\ D {\bf 84}, 061702 (2011).
%  [arXiv:1103.3829 [hep-th]].
  %%CITATION = ARXIV:1103.3829;%%


%%%%%%%%%%%%
%  interplay lattice and SDE
%%%%%%%%%%%%%

%\cite{Aguilar:2010zx}
\bibitem{Aguilar:2010zx} 
  A.~C.~Aguilar, D.~Binosi and J.~Papavassiliou,
  %``Nonperturbative gluon and ghost propagators for d=3 Yang-Mills,''
  Phys.\ Rev.\ D {\bf 81}, 125025 (2010).
%  [arXiv:1004.2011 [hep-ph]].
  %%CITATION = ARXIV:1004.2011;%%
  
  %\cite{Aguilar:2010gm}
\bibitem{Aguilar:2010gm} 
  A.~C.~Aguilar, D.~Binosi and J.~Papavassiliou,
  %``QCD effective charges from lattice data,''
  JHEP {\bf 1007}, 002 (2010).
%  [arXiv:1004.1105 [hep-ph]].
  %%CITATION = ARXIV:1004.1105;%%
  

%\cite{Aguilar:2010cn}
\bibitem{Aguilar:2010cn}
A.~C.~Aguilar and J.~Papavassiliou,
%``Chiral symmetry breaking with lattice propagators,''
Phys.\ Rev.\ D {\bf 83}, 014013 (2011).
%  [arXiv:1010.5815 [hep-ph]].
%%CITATION = PHRVA,D83,014013;%%



%\cite{Aguilar:2011yb}
\bibitem{Aguilar:2011yb} 
  A.~C.~Aguilar, D.~Binosi and J.~Papavassiliou,
  %``Gluon mass through ghost synergy,''
  JHEP {\bf 1201}, 050 (2012).
%  [arXiv:1108.5989 [hep-ph]].
  %%CITATION = ARXIV:1108.5989;%%


%\cite{Cucchieri:2011ig}
\bibitem{Cucchieri:2011ig} 
  A.~Cucchieri, D.~Dudal, T.~Mendes and N.~Vandersickel,
  %``Modeling the Gluon Propagator in Landau Gauge: Lattice Estimates of Pole Masses and Dimension-Two Condensates,''
  arXiv:1111.2327 [hep-lat].
  %%CITATION = ARXIV:1111.2327;%%


%\cite{Dudal:2012hb}
\bibitem{Dudal:2012hb} 
  D.~Dudal, N.~Vandersickel, A.~Cucchieri and T.~Mendes,
  %``Ghost dissection,''
  PoS QCD {\bf -TNT-II}, 015 (2011).
%  [arXiv:1202.2208 [hep-th]].
  %%CITATION = ARXIV:1202.2208;%%

  
%\cite{Skullerud:2003qu}
\bibitem{Skullerud:2003qu} 
  J.~I.~Skullerud, P.~O.~Bowman, A.~Kizilersu, D.~B.~Leinweber and A.~G.~Williams,
  %``Nonperturbative structure of the quark gluon vertex,''
  JHEP {\bf 0304}, 047 (2003).
%  [hep-ph/0303176].
  %%CITATION = HEP-PH/0303176;%%

%\cite{Cucchieri:2008qm}
\bibitem{Cucchieri:2008qm} 
  A.~Cucchieri, A.~Maas and T.~Mendes,
  %``Three-point vertices in Landau-gauge Yang-Mills theory,''
  Phys.\ Rev.\ D {\bf 77}, 094510 (2008).
% [arXiv:0803.1798 [hep-lat]].
  %%CITATION = ARXIV:0803.1798;%%
    
    
%\cite{Boucaud:2011eh}
\bibitem{Boucaud:2011eh} 
  P.~.Boucaud, D.~Dudal, J.~P.~Leroy, O.~Pene and J.~Rodriguez-Quintero,
  %``On the leading OPE corrections to the ghost-gluon vertex and the Taylor theorem,''
  JHEP {\bf 1112}, 018 (2011).
%  [arXiv:1109.3803 [hep-ph]].
  %%CITATION = ARXIV:1109.3803;%%    

%%%%%%%%%%%%%%%%%5
%  previous studies
%%%%%%%%%%%%%%%%%%


%\cite{Fischer:2003rp}
\bibitem{Fischer:2003rp} 
  C.~S.~Fischer and R.~Alkofer,
  %``Nonperturbative propagators, running coupling and dynamical quark mass of Landau gauge QCD,''
  Phys.\ Rev.\ D {\bf 67}, 094020 (2003).
%  [hep-ph/0301094].
  %%CITATION = HEP-PH/0301094;%%


%\cite{Fischer:2005en}
\bibitem{Fischer:2005en} 
  C.~S.~Fischer, P.~Watson and W.~Cassing,
  %``Probing unquenching effects in the gluon polarisation in light mesons,''
  Phys.\ Rev.\ D {\bf 72}, 094025 (2005).
  %  [hep-ph/0509213].
  %%CITATION = HEP-PH/0509213;%%
  


%\cite{Cornwall:1981zr}
\bibitem{Cornwall:1981zr}
J.~M.~Cornwall,
%``Dynamical Mass Generation In Continuum QCD,''
Phys.\ Rev.\ D {\bf 26}, 1453 (1982).
%%CITATION = PHRVA,D26,1453;%%

%\cite{Cornwall:1989gv}
\bibitem{Cornwall:1989gv}
J.~M.~Cornwall and J.~Papavassiliou,
%``Gauge Invariant Three Gluon Vertex in QCD,''
Phys.\ Rev.\  D {\bf 40}, 3474 (1989).
%%CITATION = PHRVA,D40,3474;%%



%\cite{Binosi:2002ft}
\bibitem{Binosi:2002ft}
D.~Binosi and J.~Papavassiliou,
%``The pinch technique to all orders,''
Phys.\ Rev.\  D {\bf 66}(R), 111901 (2002).
%[arXiv:hep-ph/0208189].
%%CITATION = PHRVA,D66,111901;%%

%\cite{Binosi:2003rr}
\bibitem{Binosi:2003rr}
D.~Binosi and J.~Papavassiliou,
%``Pinch technique self-energies and vertices to all orders in perturbation
%theory,''
J.\ Phys.\ G {\bf 30}, 203 (2004).
%[arXiv:hep-ph/0301096].
%%CITATION = JPHGB,G30,203;%%


%\cite{Abbott:1980hw}
\bibitem{Abbott:1980hw}
See, e.g.,  L.~F.~Abbott,
%``The Background Field Method Beyond One Loop,''
Nucl.\ Phys.\  B {\bf 185}, 189 (1981), and references therein.
%%CITATION = NUPHA,B185,189;%%


%\cite{Binosi:2008qk}
\bibitem{Binosi:2008qk}
D.~Binosi and J.~Papavassiliou,
%``New Schwinger-Dyson equations for non-Abelian gauge theories,''
JHEP {\bf 0811}, 063 (2008).
%[arXiv:0805.3994 [hep-ph]].
%%CITATION = JHEPA,0811,063;%%



%\cite{Marciano:1977su}
\bibitem{Marciano:1977su} 
  W.~J.~Marciano and H.~Pagels,
  %``Quantum Chromodynamics: A Review,''
  Phys.\ Rept.\  {\bf 36}, 137 (1978).
  %%CITATION = PRPLC,36,137;%%
  


%\cite{Ball:1980ay}
\bibitem{Ball:1980ay}
  J.~S.~Ball, T.~-W.~Chiu,
  %``Analytic Properties of the Vertex Function in Gauge Theories. 1.,''
  Phys.\ Rev.\  {\bf D22}, 2542 (1980).
  

%\cite{Curtis:1990zs}
\bibitem{Curtis:1990zs}
  D.~C.~Curtis and M.~R.~Pennington,
  %``Truncating the Schwinger-Dyson equations: How multiplicative
  %renormalizability and the Ward identity restrict the three point vertex in
  %QED,''
  Phys.\ Rev.\  D {\bf 42}, 4165 (1990).
  %%CITATION = PHRVA,D42,4165;%%
  



%\cite{Grassi:2004yq}
\bibitem{Grassi:2004yq}
  P.~A.~Grassi, T.~Hurth and A.~Quadri,
  %``On the Landau background gauge fixing and the IR properties of YM Green
  %functions,''
  Phys.\ Rev.\  D {\bf 70}, 105014 (2004).
%  [arXiv:hep-th/0405104].
  %%CITATION = PHRVA,D70,105014;%%


%\cite{Aguilar:2009pp}
\bibitem{Aguilar:2009pp}
  A.~C.~Aguilar, D.~Binosi and J.~Papavassiliou,
  %``Indirect determination of the Kugo-Ojima function from lattice data,''
  JHEP {\bf 0911}, 066 (2009).
 % [arXiv:0907.0153 [hep-ph]].
  %%CITATION = JHEPA,0911,066;%%
    

%\cite{Aguilar:2009nf}
\bibitem{Aguilar:2009nf}
  A.~C.~Aguilar, D.~Binosi, J.~Papavassiliou and J.~Rodriguez-Quintero,
  %``Non-perturbative comparison of QCD effective charges,''
  Phys.\ Rev.\  D {\bf 80}, 085018 (2009).
%  [arXiv:0906.2633 [hep-ph]].
  %%CITATION = PHRVA,D80,085018;%%




%\cite{Sternbeck:2006rd}
\bibitem{Sternbeck:2006rd} 
  A.~Sternbeck,
  %``The Infrared behavior of lattice QCD Green's functions,''
  hep-lat/0609016.
  %%CITATION = HEP-LAT/0609016;%%
 
  
%\cite{Aguilar:2011ux}
\bibitem{Aguilar:2011ux} 
  A.~C.~Aguilar, D.~Binosi and J.~Papavassiliou,
  %``The dynamical equation of the effective gluon mass,''
  Phys.\ Rev.\ D {\bf 84}, 085026 (2011).
%  [arXiv:1107.3968 [hep-ph]].
  %%CITATION = ARXIV:1107.3968;%%


%\cite{Aguilar:2009ke}
\bibitem{Aguilar:2009ke}
  A.~C.~Aguilar and J.~Papavassiliou,
  %``Gluon mass generation without seagull divergences,''
  Phys.\ Rev.\  D {\bf 81}, 034003 (2010).
 % [arXiv:0910.4142 [hep-ph]].
  %%CITATION = PHRVA,D81,034003;%%



%\cite{Aguilar:2011xe}
\bibitem{Aguilar:2011xe} 
  A.~C.~Aguilar, D.~Ibanez, V.~Mathieu and J.~Papavassiliou,
  %``Massless bound-state excitations and the Schwinger mechanism in QCD,''
  Phys.\ Rev.\ D {\bf 85}, 014018 (2012).
%  [arXiv:1110.2633 [hep-ph]].
  %%CITATION = ARXIV:1110.2633;%%  



%\cite{DelDebbio:2010zz}
\bibitem{DelDebbio:2010zz} 
  L.~Del Debbio,
  %``The conformal window on the lattice,''
  PoS LATTICE {\bf 2010}, 004 (2010).
  %%CITATION = POSCI,LATTICE2010,004;%%


%\cite{Cheng:2011qc}
\bibitem{Cheng:2011qc} 
  X.~Cheng and E.~T.~Tomboulis,
  %``Fermion RG blocking transformations and IR structure,''
  PoS QCD {\bf -TNT-II}, 046 (2011).
%  [arXiv:1112.4235 [hep-lat]].
  %%CITATION = ARXIV:1112.4235;%%
  

%\cite{Davydychev:2000rt}
\bibitem{Davydychev:2000rt}
  A.~I.~Davydychev, P.~Osland and L.~Saks,
  %``Quark gluon vertex in arbitrary gauge and dimension,''
  Phys.\ Rev.\  D {\bf 63}, 014022 (2001).
% [arXiv:hep-ph/0008171].
  %%CITATION = PHRVA,D63,014022;%%    



%\cite{Roberts:1994dr}
\bibitem{Roberts:1994dr}
  C.~D.~Roberts and A.~G.~Williams,
  %``Dyson-Schwinger equations and their application to hadronic physics,''
  Prog.\ Part.\ Nucl.\ Phys.\  {\bf 33}, 477 (1994).
%  [arXiv:hep-ph/9403224].
  %%CITATION = PPNPD,33,477;%%



%\cite{Grassi:1999tp}
\bibitem{Grassi:1999tp} 
  P.~A.~Grassi, T.~Hurth and M.~Steinhauser,
  %``Practical algebraic renormalization,''
  Annals Phys.\  {\bf 288}, 197 (2001).
%  [hep-ph/9907426].
  %%CITATION = HEP-PH/9907426;%%


  



%\cite{Maris:1997tm}
\bibitem{Maris:1997tm} 
  P.~Maris and C.~D.~Roberts,
  %``Pi- and K meson Bethe-Salpeter amplitudes,''
  Phys.\ Rev.\ C {\bf 56}, 3369 (1997).
%  [nucl-th/9708029].
  %%CITATION = NUCL-TH/9708029;%%

%\cite{ElBennich:2011py}
\bibitem{ElBennich:2011py} 
  B.~El-Bennich, G.~Krein, L.~Chang, C.~D.~Roberts and D.~J.~Wilson,
  %``Flavor SU(4) breaking between effective couplings,''
  Phys.\ Rev.\ D {\bf 85}, 031502 (2012).
%  [arXiv:1111.3647 [nucl-th]].
  %%CITATION = ARXIV:1111.3647;%%
  
%\cite{Kizilersu:2009kg}
\bibitem{Kizilersu:2009kg}
  A.~Kizilersu and M.~R.~Pennington,
  %``Building the Full Fermion-Photon Vertex of QED by Imposing Multiplicative
  %Renormalizability of the Schwinger-Dyson Equations for the Fermion and Photon
  %Propagators,''
  Phys.\ Rev.\  D {\bf 79}, 125020 (2009);
 % [arXiv:0904.3483 [hep-th]].
  %%CITATION = PHRVA,D79,125020;%%
%\cite{Bashir:1997qt}
%\bibitem{Bashir:1997qt}
  A.~Bashir, A.~Kizilersu and M.~R.~Pennington,
  %``The non-perturbative three-point vertex in massless quenched QED and
  %perturbation theory constraints,''
  Phys.\ Rev.\  D {\bf 57}, 1242 (1998).
%  [arXiv:hep-ph/9707421].
  %%CITATION = PHRVA,D57,1242;%%  



%\cite{Bashir:2011dp}
\bibitem{Bashir:2011dp}
  A.~Bashir, R.~Bermudez, L.~Chang and C.~D.~Roberts,
  %``Dynamical chiral symmetry breaking and the fermion--gauge-boson vertex,''
  arXiv:1112.4847 [nucl-th].
  %%CITATION = ARXIV:1112.4847;%%


  


\end{thebibliography}
\end{document}